\documentclass[12pt]{article}

\usepackage[lmargin=1.2in,rmargin=1.2in]{geometry}
\usepackage{amsmath, amssymb, amsfonts, amscd, xspace, pifont}

\usepackage[]{algorithm2e}
\usepackage{graphicx}
\usepackage{hhline} 
\usepackage{multirow}
\usepackage{url}
\usepackage{enumerate}

\usepackage{bm}

\newtheorem{lemma}{Lemma}
\newtheorem{corollary}{Corollary}

\newcommand{\diag}{\text{diag}}
\newcommand{\var}{\text{var}}
\newcommand{\cov}{\text{cov}}
\newcommand{\rank}{\text{rank}}

\newcommand{\db}{\mathbf{d}}
\newcommand{\eb}{\mathbf{e}}

\newcommand{\qb}{\mathbf{q}}

\newcommand{\wb}{\mathbf{w}}
\newcommand{\xb}{\mathbf{x}}

\newcommand{\zb}{\mathbf{z}}

\newcommand{\Ab}{\mathbf{A}}
\newcommand{\Bb}{\mathbf{B}}
\newcommand{\Cb}{\mathbf{C}}
\newcommand{\Db}{\mathbf{D}}
\newcommand{\Eb}{\mathbf{E}}

\newcommand{\Ib}{\mathbf{I}}
\newcommand{\Jb}{\mathbf{J}}

\newcommand{\Mb}{\mathbf{M}}

\newcommand{\Pb}{\mathbf{P}}
\newcommand{\Qb}{\mathbf{Q}}
\newcommand{\Rb}{\mathbf{R}}
\newcommand{\Sbb}{\mathbf{S}}
\newcommand{\Tb}{\mathbf{T}}
\newcommand{\Ub}{\mathbf{U}}
\newcommand{\Vb}{\mathbf{V}}
\newcommand{\Wb}{\mathbf{W}}
\newcommand{\Xb}{\mathbf{X}}
\newcommand{\Yb}{\mathbf{Y}}
\newcommand{\Zb}{\mathbf{Z}}

\newcommand{\bA}{\bm{A}}

\newcommand{\bJ}{\bm{J}}

\newcommand{\bS}{\bm{S}}
\newcommand{\bT}{\bm{T}}

\newcommand{\bY}{\bm{Y}}

\newcommand{\alphab}{\bm{\alpha}}

\newcommand{\Lambdab}{\bm{\Lambda}}

\newcommand{\bbR}{\mathbb{R}}

\newcommand{\Ug}{\Ub_g}

\newcommand{\Xg}{\Xb_g}

\newcommand{\Yg}{\bY_g}

\newcommand{\Xindv}{\Xb_g^{\text{Indiv}}}

\newcommand{\Yindv}{\bY_g^{\text{Indiv}}}
\newcommand{\Xjoint}{\Xb^{\text{Joint}}}
\newcommand{\Xgjoint}{\Xg^{\text{Joint}}}
\newcommand{\Yjoint}{\bY^{\text{Joint}}}

\newcommand{\tWb}{\tilde{\Wb}}
\newcommand{\row}{\text{row}}
\newcommand{\col}{\text{col}}
\usepackage{xcolor}
\usepackage{xr}
\usepackage{natbib}
\usepackage{setspace}
\numberwithin{equation}{section}
\numberwithin{figure}{section}

\usepackage[figuresright]{rotating}

\begin{document}
\def\spacingset#1{\renewcommand{\baselinestretch}%
{#1}\small\normalsize} \spacingset{1.5}
\title{\bf Joint and Individual Component Regression}
    
\author{
	\vspace{-0.3cm}
	Peiyao Wang$^{1}$, 
	Haodong Wang$^{1}$, 
	Quefeng Li$^{2}$, 
	Dinggang Shen$^{4, 5, 6}$,\\ 
	\vspace{-0.7cm}
	Yufeng Liu$^{1,2,3*}$ \\ \\ 
	\vspace{-0.3cm}
	$^1$Department of Statistics and Operations Research, \\
	\vspace{-0.3cm}
	$^2$Department of Biostatistics, 
	$^3$Department of Genetics, \\
	\vspace{-0.3cm}
	University of North Carolina at Chapel Hill; \\ 
	\vspace{-0.3cm}
	$^4$School of Biomedical Engineering, ShanghaiTech University; \\ 
	\vspace{-0.3cm}
	$^5$Shanghai United Imaging Intelligence Co., Ltd.;\\ 
	\vspace{-0.3cm}
	$^6$Department of Artificial Intelligence, Korea University \\
	$^*$\emph{email}: yfliu@email.unc.edu}

\maketitle








\label{firstpage}


\begin{abstract}
  Multi-group data are commonly seen in practice. Such data structure consists of data from multiple groups and can be challenging to analyze due to data heterogeneity. We propose a novel Joint and Individual Component Regression (JICO) model to analyze multi-group data. In
particular, our proposed model decomposes the response into shared and group-specific components, which are driven by
low-rank approximations of joint and individual structures from the predictors respectively. The joint structure has the
same regression coefficients across multiple groups, whereas individual structures have group-specific regression
coefficients. Moreover, the choice of global and individual ranks allows our model to cover global and group-specific
models as special cases. For model estimation, we formulate this framework under the representation of latent components
and propose an iterative algorithm to solve for the joint and individual scores under the new representation. To
construct the latent scores, we utilize the Continuum Regression (CR), which provides a unified framework that covers
the Ordinary Least Squares (OLS), the Partial Least Squares (PLS), and the Principal Component Regression (PCR) as its
special cases. We show that JICO attains a good balance between global and group-specific models and remains flexible
due to the usage of CR. Finally, we conduct simulation studies and  analysis of an Alzheimer's disease  dataset to further
demonstrate the effectiveness of JICO. 
R implementation of JICO is available online at \url{https://github.com/peiyaow/JICO}.
\end{abstract}

%




%

\section{Introduction}
Many fields of scientific research involve the analysis of heterogeneous data. In particular, data may appear in the form
of multiple matrices, with data heterogeneity arising from either variables or
samples. 
One example is the multi-view/source data, 
 which measure different sets of variables on the same set
of samples. The sets of variables may come from different platforms/sources/modalities. For instance, in genomics studies,
measurements are collected as different biomarkers, such as mRNA and miRNA \citep{muniategui2013joint}. 
Another example is the multi-group data, 
 which measure the same set of variables on disparate
sets of samples, which leads to heterogenous subpopulations/subgroups in the entire population. For instance, in the
Alzheimer's Disease (AD) study, subjects can have different subtypes, such as Normal Control (NC), Mild Cognitive
Impairment (MCI), and AD. 

We 
study the classical regression problem with one continuous response for multi-group data. Although there are many
well-established regression techniques for homogeneous data \citep{hoerl1970ridge, tibshirani1996regression}, 
they may not be suitable for multi-group data.
One naive approach is to ignore data heterogeneity and fit a global model using these techniques. However, a single global model can be too restrictive because the diverse information from different subgroups may not be identified. On the other hand, one can train separate group-specific models. Despite its flexibility, the information that is jointly shared across different groups is not sufficiently captured. Therefore, it is desirable to build a flexible statistical model that can simultaneously quantify the jointly shared global information and individual group-specific information for heterogeneous data.

There are several existing methods proposed in the literature under the context of regression for multi-group data. 
 \cite{meinshausen2015maximin} took a conservative approach and proposed a maxmin effect method that is reliable for all possible subsets of the data. \cite{zhao2016partially} proposed a partially linear regression framework for massive heterogeneous data, and the goal is to extract common features across all subpopulations while exploring heterogeneity of each subpopulation. \cite{tang2016fused, ma2017concave} proposed fused penalties to estimate regression coefficients that capture subgroup structures in a linear regression framework. \cite{wang2018flexible} proposed a locally-weighted penalized model to perform subject-wise variable selection. \cite{wang2023high} proposed a factor regression model for heterogeneous subpopulations under the high-dimensional factor decomposition. However, these models either are  not specifically designed to identify the globally-shared and group-specific structures, or impose strong theoretical assumptions on the covariates. On the other hand, there exist some works that study this manner for multi-source data.
\cite{lock2013joint} proposed JIVE to learn joint and individual structures from multiple data matrices by low-rank approximations. 
Some extensions of JIVE can be found in \cite{feng2018angle, gaynanova2019structural}. All of these decomposition methods
are fully unsupervised. Recently, \cite{li2021integrative} proposed a supervised integrative factor regression model for
mult-source data and studied its statistical properties with hypothesis tests. \cite{palzer2022sjive} proposed sJIVE that
extends JIVE with the supervision from the response. These methods are supervised, but both focused on regressions for
multi-source
data. 

In this paper, we consider the supervised learning problem of predicting a response with multi-group data. We propose a Joint and
Individual COmponent Regression (JICO), a novel latent component regression model 
that covers JIVE as a special case. 
Our proposed model decomposes the response into jointly shared and
group-specific components, which are driven by low-rank approximations of joint and individual structures from the
predictors respectively. The joint structure shares the same 
coefficients across all groups, whereas individual structures have group-specific coefficients. Moreover, by choosing
different ranks of joint and individual structures, our model covers global and group-specific models as special cases. To
estimate JICO, we 
propose an iterative algorithm to solve for joint and individual scores using latent component representation. To
construct the latent scores, we utilize the Continuum Regression (CR) \citep{stone1990continuum}, which provides a unified
framework that covers OLS, PLS, and PCR as special cases. Some follow-up studies and modern extensions of CR can be found in 
\cite{ bjorkstrom1996continuum,  lee2013kernel}. 
 Embracing this flexibility and generaliziblity from CR, our proposed JICO model extends to the heterogeneous data setup and is able to achieve different model configurations on the spectrum of CR under this more complicated  setting. 
JICO attains a good balance between global and group-specific models, and further
achieves its flexibility by extending
CR. 

The rest of this paper is organized as follows. In Section \ref{sec: jicomodel}, we briefly review JIVE and introduce
our proposed JICO model. We further provide sufficient conditions to make JICO identifiable. In Section \ref{sec: jicoestimation}, after two motivating special cases, we introduce our
iterative algorithm. In Sections \ref{sec: jicosim} and \ref{sec: jicoreal}, we evaluate the performance of JICO by
simulation studies and real data analysis on the Alzheimer's disease dataset, respectively. In Section \ref{sec: jicocon}, we conclude
this paper with some discussion and possible extensions.

\section{Motivation and Model Framework}\label{sec: jicomodel}
Suppose we observe data pairs $(\Xb_g, \bY_g)_{g = 1}^G$ from $G$ groups, where $\Xb_g \in \mathbb{R}^{n_g \times p}$ and $\Yg \in \mathbb{R}^{n_g}$ are the data matrix and the response vector for the $g$th group respectively. 
In this setting, each data matrix has the same set of $p$ explanatory variables, whereas the samples vary across groups. We let $\Xb = [\Xb_1', \ldots, \Xb_G']' \in \bbR^{n \times p}$ and $\bY = [\bY_1', \ldots, \bY_G']' \in \bbR^{n}$, where $n = \sum_{g=1}^G n_g$. 

Our model is closely related to  JIVE, which provides a general formulation to decompose multiple data matrices into joint and individual structures. 
The JIVE decomposes $\Xb_g$ as 
\begin{equation}\label{eq: jive}
\Xb_g = \Jb_g + \Ab_g + \Eb_g, 
\end{equation}
where $\Jb_g \in \bbR^{n_g \times p}$ represents the submatrix of the joint structure of $\Xg$,
$\Ab_g \in \bbR^{n_g \times p}$ represents the individual structure of $\Xg$, and $\Eb_g \in \bbR^{n_g \times p}$ is the
error matrix.  
We consider that $\bY_g$ has a similar decomposition into joint and individual signals
\begin{equation}\label{eq: Yjive}
\bY_g = \bJ^Y_g + \bA^Y_g + \eb_g, 
\end{equation}
where $\eb_g \in \bbR^{n_g}$ is the noise from the $g$-th group. Let $\tilde{\Xb}_g= \Jb_g + \Ab_g$ and
$\tilde{\bY}_g=\bJ^Y_g + \bA^Y_g$ be the noiseless counterparts of $\Xb_g$ and $\bY_g$.  Lemma 1 gives conditions to ensure that $\Jb_g$, $\Ab_g$, $\bJ_g^Y$, and $\bA_g^Y$
are identifiable. 
 
\begin{lemma}
	Given $\{\tilde{\Xb}_g, \tilde{\bY_g}\}_{g=1}^G$, where $\tilde{\bY_g} \in \col(\tilde{\Xb}_g)$. There exists unique
	$\Jb_g$ and $\Ab_g$ such that $\tilde{\Xb}_g = \Jb_g +\Ab_g $ and they satisfy the following conditions:
	\begin{enumerate}[(i)]
		\item $\row(\Jb_1) = \ldots = \row(\Jb_G) \subset \row(\tilde{\Xb}_g)$;
		\item $\row(\Jb_g) \perp \row (\Ab_g)$, for $g = 1, \ldots, G$;
		\item $\bigcap_{g=1}^G \row(\Ab_g) = \{\mathbf{0}_{p\times 1}\}$.
	\end{enumerate}
	Moreover, if $\col(\Jb_g) \perp \col(\Ab_g)$, then there exists unique $\bJ^{Y}_g$ and $\bA^{Y}_g $ such that
	$\tilde{\bY}_g = \bJ^{Y}_g +\bA^{Y}_g $ and they satisfy $\bJ^{Y}_g \in \col(\Jb_g)$ and
	$\bA^{Y}_g \in \col(\Ab_g)$.
\end{lemma}

Lemma 1 shows that $\tilde{\Xb}_g$ can be uniquely decomposed into the sum of $\Jb_g$ and $\Ab_g$ if we require them to
satisfy conditions $(i) - (iii)$, following similar statements as in \cite{feng2018angle}. To ensure the unique decomposition of $\tilde{\Yb}_g$, we need to further  require
$\col(\Jb_g) \perp \col(\Ab_g)$, which is different from \cite{palzer2022sjive}, that requires $\row(\Jb_g) \perp \row(\Ab_g)$.

In practice, only $\Xb_g$ and $\bY_g$ are observable. In Lemma 2, we show in $(a)$ that the identifiable conditions in Lemma
1 can still be achieved given the observed $\{\Xb_g, \bY_g\}_{g=1}^G$.
Moreover, in Lemma 2$(b)$, we show that if $\Jb_g$ and $\Ab_g$ are assumed to have low ranks, 
they can be further decomposed
as 
$\Jb_g = \Sbb_g \Ub$, where $\Sbb_g$ is a $n_g \times K$ score matrix, $\Ub$ is a $K \times p$ loading matrix, and
$K=\rank(\Jb_g)$; and $\Ab_g = \Tb_g \Ub_g$, where $\Tb_g$ is a $n_g \times K_g$ score matrix and $\Ub_g$ is a
$K_g \times p$ loading matrix, and $K_g=\rank(\Ab_g)$. Under this formulation, if $\Sbb_g'\Tb_g = \bm{0}_{K\times K_g}$, then 
(\ref{eq: jive}) and (\ref{eq: Yjive}) can be expressed as 
\begin{gather}
\Xb_g = \Sbb_g \Ub + \Tb_g\Ug + \Eb_g, \label{eq: Xlatentdecomp}\\
\bY_g  = \Sbb_g \alphab  + \Tb_g\alphab_g + \eb_g, \label{eq: Ylatentdecomp}
\end{gather}
where $\alphab \in \bbR^K$ and $\alphab_g \in \bbR^{K_g}$ are the coefficients of the joint and individual components
respectively. Model \eqref{eq: Ylatentdecomp} gives a unified framework to model multi-group data. 
When $K = 0$, the joint term $\Sbb_g \alphab$ vanishes and \eqref{eq: Ylatentdecomp} reduces to a group-specific model of
$\bY_g = \Tb_g\alphab_g + \eb_g$. On the other hand, when $K_1 =\cdots=K_g= 0$, the individual term $\Tb_g\alphab_g$
vanishes and \eqref{eq: Ylatentdecomp} reduces to a global model of $\bY_g = \Sbb_g\alphab + \eb_g$. When $K \ne 0$ and
$K_g \ne 0$, \eqref{eq: Ylatentdecomp} has both global and group-specific components, thus lies between the above two extreme cases.

\begin{lemma}Given $\{{\Xb}_g, {\bY_g}\}_{g=1}^G$, it holds that 
	\begin{enumerate}[(a)]
		\item There exists matrices $\Jb_g$ and $\Ab_g$ such that they satisfy conditions $(i) - (iii)$ in Lemma 1 and  $\col(\Jb_g) \perp \col(\Ab_g)$.  
		\item There exists matrices $\Ub, \Ub_g, \Sbb_g, \Tb_g$ such that $\Jb_g$ and $\Ab_g$ can be expressed as
		$\Jb_g = \Sbb_g \Ub$ and $\Ab_g = \Tb_g \Ub_g$, where $\Sbb_g'\Tb_g = \bm{0}_{K\times K_g}$.  Moreover, there exists $\bJ^{Y}_g$ and $\bA^{Y}_g$ which can be expressed as 
		$\bJ^{Y}_g = \Sbb_g\alphab$ and 
		$\bA^{Y}_g = \Tb_g \alphab_g$, where 
		${\alphab} = ({\Sbb}'{\Sbb})^{-1} {\Sbb}'\bY$ and ${\alphab}_g = ({\Tb}_g'{\Tb}_g)^{-1}{\Tb}_g'{\bY}_g$, with $\Sbb = [\Sbb_1', \ldots, \Sbb_G']$. 
	\end{enumerate}

\end{lemma}


\begin{corollary}
	There exist matrices $\Wb \in \mathbb{R}^{p \times K}$ and ${\Wb_g} \in \mathbb{R}^{p \times K_g}$ such that $\Jb_g = \Sbb_g \Ub$ and $\Ab_g = \Tb_g \Ub_g$ 
	 defined by $\Sbb_g = {\Xb}_g \Wb $, $\Tb_g = {\Xb}_g \Wb_g$, $\Ub = (\Wb'\Wb)^{-1} \Wb'$ and
	$\Ub_g = (\Wb_g'\Wb_g)^{-1} \Wb_g'$ as in Lemma 2$(b)$ satisfy conditions $(i) - (iii)$ in Lemma 1 and  $\col(\Jb_g) \perp \col(\Ab_g)$, if $\Wb'\Wb_g = \bm{0}$ and $\Wb'{\Xb}_g'{\Xb}_g\Wb_g = \bm{0}$, for all $g$.
\end{corollary}

Corollary 1 follows directly from the proof of Lemma 2$(b)$. As a remark, the columns of $\Wb$ and $\Wb_g$ form the sets of bases that span the row spaces of $\Jb_g$ and $\Ab_g$ respectively. Hence, $\Wb'\Wb_g = \bm{0}$ is a sufficient and necessary condition for $\row(\Jb_g) \perp \row(\Ab_g)$. Moreover, note that in Lemma 2$(b)$, $\Sbb_g'\Tb_g = \bm{0}$ directly implies that $\Ab_g' \Jb_g = \bm{0}$, the latter being a sufficient condition for $\col(\Jb_g) \perp \col (\Ab_g)$. Therefore, in Corollary 1, $\Wb'{\Xb}_g'{\Xb}_g\Wb_g = \bm{0}$ provides a sufficient condition for $\col(\Jb_g) \perp \col (\Ab_g)$, which satisfies one of the identifiability constraints for the unique decomposition of $\Xb_g$ in Lemma 1. In Section \ref{sec: jicoestimation}, we describe the algorithm to solve for $\Wb$ and $\Wb_g$ respectively.

	
\section{Model Estimation}\label{sec: jicoestimation}
The key to estimate \eqref{eq: Xlatentdecomp} and \eqref{eq: Ylatentdecomp} is 
the constructions of score matrices $\Sbb_g$ and $\Tb_g$. To motivate our estimation procedure, in Sections \ref{sec: joint
	score}, we discuss the joint and individual score estimation under two special
cases respectively. In Section \ref{sec: iterative}, we introduce an iterative algorithm for the general case.  


\subsection{Joint and Individual Score Estimation}\label{sec: joint score}
We first consider a 
special case that $K_g = 0, g = 1, \ldots, G$. Under this setup, the individual components vanish and 
\eqref{eq: Xlatentdecomp} and \eqref{eq: Ylatentdecomp} reduce to the following model:
\begin{gather}
\Xb = \Sbb \Ub + \Eb,\quad  \bY = \Sbb \alphab + \eb, \label{eq: latentjoint}
\end{gather}
where $\Sbb = [\Sbb_1', \ldots, \Sbb_G']'$, $\Eb = [\Eb_1', \ldots, \Eb_G]'$, and  $\eb = [\eb_1', \ldots, \eb_G']'$. 

The formulation of \eqref{eq: latentjoint} covers many existing classic methods. For example, in PCR, $\Sbb$ is chosen to be the score matrix of
the first $K$ principal components of $\Xb'\Xb$. However, PCR is inherently unsupervised and ignores the information from $\bY$. Among the other supervised methods, PLS regression is a popular approach that incorporates regression on the latent scores. When $K = 1$ and $p < n$, the standard OLS can also be cast under the above setup.


According to the proof of our Lemma 2, $\Sbb$ can be constructed with the basis matrix $\Wb$.  For the estimation of
$\Wb$, we utilize the continuum regression (CR) \citep{stone1990continuum} algorithm,
the result of which covers OLS, PLS, and PCR as special cases. 
For $k = 1, \ldots, K$, CR sequentially solves $\wb_k$ from the following optimization
problem:
\begin{equation}\label{eq: continuumopt}
\begin{aligned}
\max_\wb\quad
&\cov(\Xb\wb, \bY)^2 \var(\Xb\wb)^{\gamma-1}\\
\text{s.t.} \quad&\wb' \wb = 1, 
\wb' \Xb'\Xb \wb_j =0; \quad   j = 1, \ldots, k -1 \text{ if }k\ge 2,
\end{aligned}
\end{equation}
where $\cov(\Xb\wb, \bY) = \wb'\Xb'\bY$ and $\var(\Xb\wb) = \wb'\Xb'\Xb \wb$, once columns of $\Xb$ and $\bY$ are
centralized to have mean zero.  Here, $\gamma \ge 0$ is a tuning parameter that controls how much variability of $\Xb$ is
taken into account for the construction of $\Sbb$. When $\gamma \rightarrow \infty$, the objective function in \eqref{eq:
	continuumopt} is dominated by $\var(\Xb\wb)^{\gamma - 1}$ and $\bY$ does not play a role. The CR solution
of $\Wb$ then seeks to find the principal component directions that maximize the variation of $\Xb$. It can be shown
that \eqref{eq: continuumopt} coincides with OLS and PLS solutions when $\gamma = 0$ and $1$ respectively. 

Let $\hat{\Wb}$ denote the solution to \eqref{eq: continuumopt} and $\hat{\Sbb} = \Xb \hat{\Wb}$. Then $\alphab$ can be
estimated by 
$\hat{\alphab} = (\hat{\Sbb}'\hat{\Sbb})^{-1} \hat{\Sbb}'\bY.$ As illustrated in Lemma 2, $\Jb_g$ is the projection of $\Xb_g$
onto the column space spanned by $\Wb$. Hence, we have $\hat{\Jb}_g = \Xb_g \hat{\Wb} (\hat{\Wb}'\hat{\Wb})^{-1} \hat{\Wb}'$,
which further gives $\hat{\Ub} = (\hat{\Wb}'\hat{\Wb})^{-1}\hat{\Wb}'$.

Next we consider our model estimation under the special case that $K= 0$. 
In this case, the joint component vanishes, and
\eqref{eq: Xlatentdecomp} and \eqref{eq: Ylatentdecomp} reduce to the following individual model:
\begin{gather}
\Xb_g 
= \Tb_g \Ub_g + \Eb_g, \quad \bY_g = \Tb_g \alphab_g + \eb_g. \label{eq: latentindv}
\end{gather}
Same as the above discussion for joint score estimation, we utilize CR to construct $\Tb_g = \Xb_g \Wb_g$ as linear transformation of $\Xb_g$, where $\Wb_g$ is a $p \times K_g$ basis matrix, whose columns span $\row(\Ab_g)$. Let $\Wb_g = [\wb_{g1}, \ldots, \wb_{g K_g}]$. Given group $g$, for $k = 1, \ldots, K_g$, CR sequentially solves $\wb_{gk}$ from the following optimization problem:
\begin{equation}\label{eq: continuumoptgroup}
\begin{aligned}
\max_\wb\quad
&\cov(\Xg\wb, \bY_g)^2 \var(\Xg\wb)^{\gamma-1}\\
\text{s.t.} \quad&\wb' \wb = 1, \\
&\wb' \Xg'\Xg\wb_{gj} =0; \quad   j = 1, \ldots, k -1 \text{ if } k\ge 2.
\end{aligned}
\end{equation}
Denote $\hat{\Wb}_g$ the solution to \eqref{eq: continuumoptgroup}. Similar to the joint estimation, once $\hat{\Tb}_g = \Xb_g \hat{\Wb}_g$ is constructed, $\alphab_g$ can be obtained as the least square solution: $\hat{\alphab}_g = (\hat{\Tb}_g'\hat{\Tb}_g)^{-1} \hat{\Tb}_g'{\bY}_g.$
Afterwards, we can have $\hat{\Ab}_g = \Xg \hat{\Wb}_g (\hat{\Wb}_g'\hat{\Wb}_g)^{-1} \hat{\Wb}_g' $ and
$\hat{\Ub}_g = (\hat{\Wb}_g'\hat{\Wb}_g)^{-1}\hat{\Wb}_g'$.

\subsection{JICO Algorithm}\label{sec: iterative}
In this section, we consider the general case where $K$ or $K_g$ can be both nonzero. Since solving \eqref{eq:
  continuumopt} and \eqref{eq: continuumoptgroup} simultaneously can be hard with both joint and individual structures
specified in the full model \eqref{eq: Xlatentdecomp} and \eqref{eq: Ylatentdecomp}, we  propose to iteratively solve
one of them while fixing the other. 
This leads to the following iterative procedure. 
\begin{itemize}
\item Given $\hat{\Wb}_g$, solve the following constrained problem sequentially for $\wb_1, \ldots, \wb_K$:
	\begin{equation}\label{eq: continuumjoint}
	\begin{aligned}
	\max_\wb\quad
	&\cov(\Xb^{\text{Joint}}\wb, \bY^{\text{Joint}})^2 \var(\Xb^{\text{Joint}}\wb)^{\gamma-1}\\
	\text{s.t.} \quad&\wb' \wb = 1, \\
	&\wb' {\Xb^{\text{Joint}}}'\Xb^{\text{Joint}} \wb_j =0; \quad   j = 1, \ldots, k -1 \text{ if }k\ge 2, \\
	&\hat{\Wb}_g'\wb = \mathbf{0}_{K_g \times 1}; \quad g = 1, \ldots, G, \\
	&\hat{\Wb}_g' {\Xb_g^{\text{Indiv}}}'\Xb_g^{\text{Joint}} \wb=\mathbf{0}_{K_g \times 1}; \quad g = 1, \ldots, G. \\
	\end{aligned}
	\end{equation}
	\item Given $\hat{\Wb}$, for any $1\leq g\leq G$, solve the following constrained problem sequentially for $\wb_{g,1}, \ldots, \wb_{g, K_g}$:
	\begin{equation}\label{eq: continuumindv}
	\begin{aligned}
	\max_\wb\quad
	&\cov(\Xindv\wb, \Yindv)^2 \var(\Xindv\wb)^{\gamma-1}\\
	\text{s.t.} \quad&\wb' \wb = 1, \\
	&\wb' {\Xindv}'\Xindv\wb_{gj} =0; \quad   j = 1, \ldots, k -1 \text{ if } k\ge 2, \\
	& \hat{\Wb}'\wb = \mathbf{0}_{K \times 1}; \\
	& \hat{\Wb}'{\Xb_g^{\text{Joint}}}'\Xb_g^{\text{Indiv}} \wb=\mathbf{0}. \\  
	\end{aligned}
	\end{equation}
	\item Repeat the above two procedures until convergence.
\end{itemize}
Note that in \eqref{eq: continuumjoint} and \eqref{eq: continuumindv}, we denote
\begin{equation*}
\Xb^{\text{Joint}} = \left[\begin{array}{c}
\Xb_1 - \Tb_1\Ub_G \\
\vdots\\
\Xb_G - \Tb_G\Ub_G
\end{array}\right], \quad
\bY^\text{Joint} = \left[\begin{array}{c}
\bY_1 - \Tb_1\alphab_1 \\
\vdots\\
\bY_G - \Tb_G\alphab_G
\end{array}\right],
\end{equation*}
and $\Xindv = \Xg - \Sbb_g \Ub, \quad \Yindv = \Yg - \Sbb_g \alphab;\quad g = 1, \ldots, G.$
Moreover, the last two constraints in \eqref{eq: continuumjoint} and \eqref{eq: continuumindv} 
correspond to the two sufficient conditions in Corollary 1 to satisfy the identifiability conditions $\row(\Jb_g) \perp \row (\Ab_g)$ and $\col(\Jb_g) \perp \col(\Ab_g)$ needed in Lemma 1.

We formulate \eqref{eq: continuumjoint} and \eqref{eq: continuumindv} into a generic CR problem, and derive an algorithm
to solve it 
in Appendix \ref{sec:Appendix B}. Furthermore, we describe the convergence criterion for the iterative procedure and give its pseudo code in
Algorithm \ref{alg: JICO} of Appendix \ref{sec:Appendix C}. Empirically, the algorithm always
meets our convergence criteria, albeit there are no theoretical guarantees. In practice, we recommend starting the algorithm with multiple initial values
and choose the one with the smallest cross-validated mean squared error. 
To predict the response using JICO estimates, we let $\Sbb_{g, test} = \Xb_{g,test} \hat{\Wb}$ and
$\Tb_{g, test} = \Xb_{g,test} \hat{\Wb}_g$, where $\Xb_{g,test}$ is the test set. Then, the prediction of response is
given by $\Sbb_{g, test} \hat{{\alphab}}+ \Tb_{g, test}\hat{{\alphab}}_g$. 

In practice,  we need to select tuning parameters $K$, $K_g$ and $\gamma$.  We describe how to select the tuning parameters in Appendix \ref{sec:Appendix D}. Since we use a  grid search for tuning parameters, if the difference between the ranks of individual matrices and joint matrix are large, our algorithm may take a long time  to compute. The goal of our algorithm is to approximate the joint and individual structure of the data by some low-rank matrices, and use the resulting low-rank matrices for prediction. Therefore, when the individual matrices and joint matrix are too complex and can not be approximated by some low-rank matrices, we suggest to use alternative methods to solve the problem.  

Finally, we point out that our method includes JIVE-predict \citep{kaplan2017prediction} as a special case. JIVE-predict
is a two-stage method that implements JIVE on $\mathbf{X}$ first and then regresses the responses on the loading
matrix. When we let $\gamma\to\infty$ in \eqref{eq: continuumopt} and \eqref{eq: continuumoptgroup}, JICO is equivalent as
performing JIVE on $\mathbf X$. For that reason, our method in that case is equivalent to JIVE-predict.

\section{Simulation Studies}\label{sec: jicosim}
One significant advantage of our proposed model is its flexibility of lying in between global and group-specific models. Moreover, the choice of parameter $\gamma$ in CR allows it to identify the model that best fits the data. In this section, we conduct multiple simulation studies to further demonstrate the advantage of our proposed model. 

We consider three simulation settings in this section. In the first two settings, we generate data according to models
that contain both global and group-specific components. The data are generated in a way that PCR and PLS solutions are
favored respectively. In the last setting, we simulate data from two special cases: a global model and a group-specific
model. The data are simulated so that the OLS is favored for both cases. For all three settings, JICO can adaptively
choose the correct model parameter $\gamma$ so that it has the optimal performance. Moreover, we further illustrate how the
rank selection impacts the performance of JICO by examining the results using mis-specified ranks.

We fix $G = 2, p = 200, n_1 = n_2 = 50$. In each replication, we generate 100 training samples to train the models and evaluate the corresponding Mean Squared Error (MSE) in an independent test set of 100 samples. We repeat simulations for 50 times. 

For $g = 1, \ldots, G$, we generate $\Xg$ as i.i.d. samples from $\mathcal{N}(\bm{0}, \Ib_{p \times p})$. For the sake of
simplicity, we generate $\Yg$ by the following model with two latent components:
\begin{equation}\label{eq: Ylatentdecompsim}
\Yg = \alpha\bS_g  + \alpha_g\bT_g  +\eb_g, 
\end{equation}
where $\bS_g = \Xb_g \wb \in \mathbb{R}^{n_g}$ is the joint latent score vector with an coefficient $\alpha$, $\bT_g =
\Xb_g \wb_g \in \mathbb{R}^{n_g}$ is the individual latent score vector with an coefficient $\alpha_g$, and $\eb_g$ is generated as i.i.d. samples from $\mathcal{N}(0, 0.04)$. Here, $\wb$ and $\wb_g$ are all $p \times 1$ vectors, and they are constructed such that $\wb'\wb_g = 0$. We vary the choices of $\wb$, $\wb_g$, $\alpha$, and $\alpha_g$, which will be discussed later. 
\subsection{PCR Setting}\label{sec: PCRsim}
In this section, we simulate the model which favors $\gamma = \infty$. In this case, CR solutions to \eqref{eq: continuumopt} and \eqref{eq: continuumoptgroup} coincide with PCR, which are essentially the top eigenvectors of the corresponding covariance matrices. 

To simulate this setup, given training data $\Xb = [\Xb_1', \Xb_2']'$, we let $\wb$ be the top eigenvector corresponding
to $\Xb'\Xb$. We further set $\wb_g$ as the top eigenvector of $\tilde{\Xb}_g' \tilde{\Xb}_g$, where $\tilde{\Xb}_g = \Xb_g (\Ib - \wb \wb')$ is the data matrix after projecting $\Xb_g$ into the linear subspace that is orthogonal to $\wb$. This projection ensures that the construction of $\wb$ and $\wb_g$ satisfies $\wb'\wb_g = 0$. To generate $\bY_g$, we let $\alpha = 1, \alpha_g = 1, g = 1, 2$.

\begin{figure}
	\begin{minipage}[b]{.5\textwidth}
		\centering
		\includegraphics[scale=0.2]{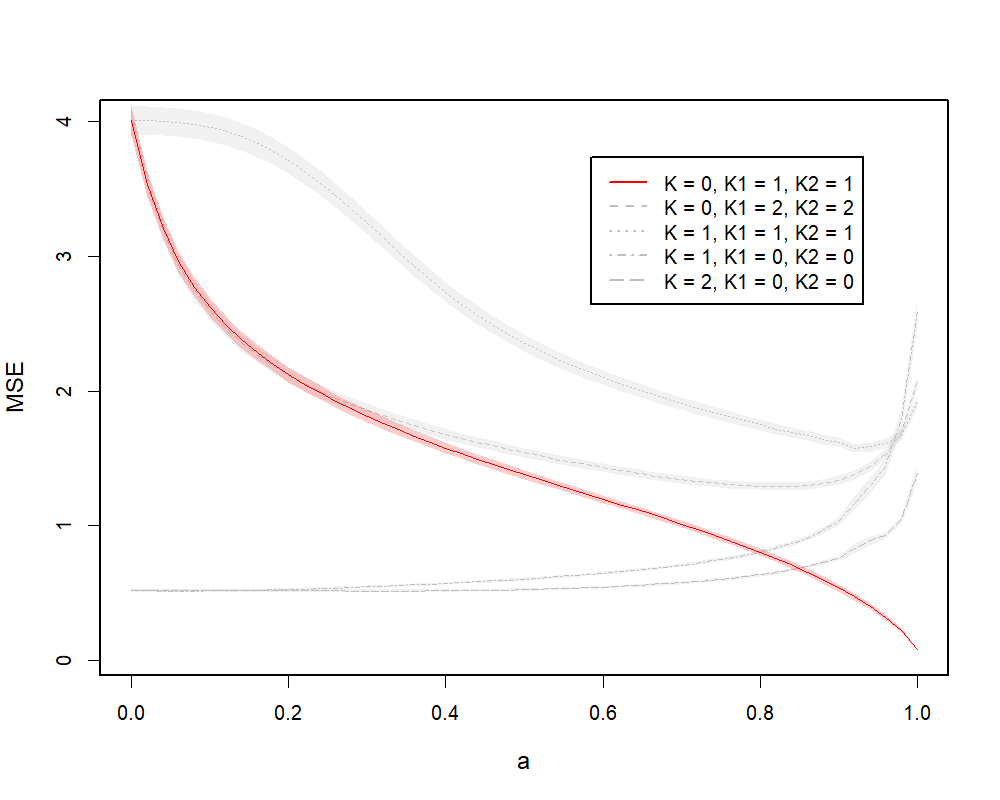}
		
		{(a) PCR Example}
	\end{minipage}
	\begin{minipage}[b]{.5\textwidth}
		\centering
		\includegraphics[scale=0.2]{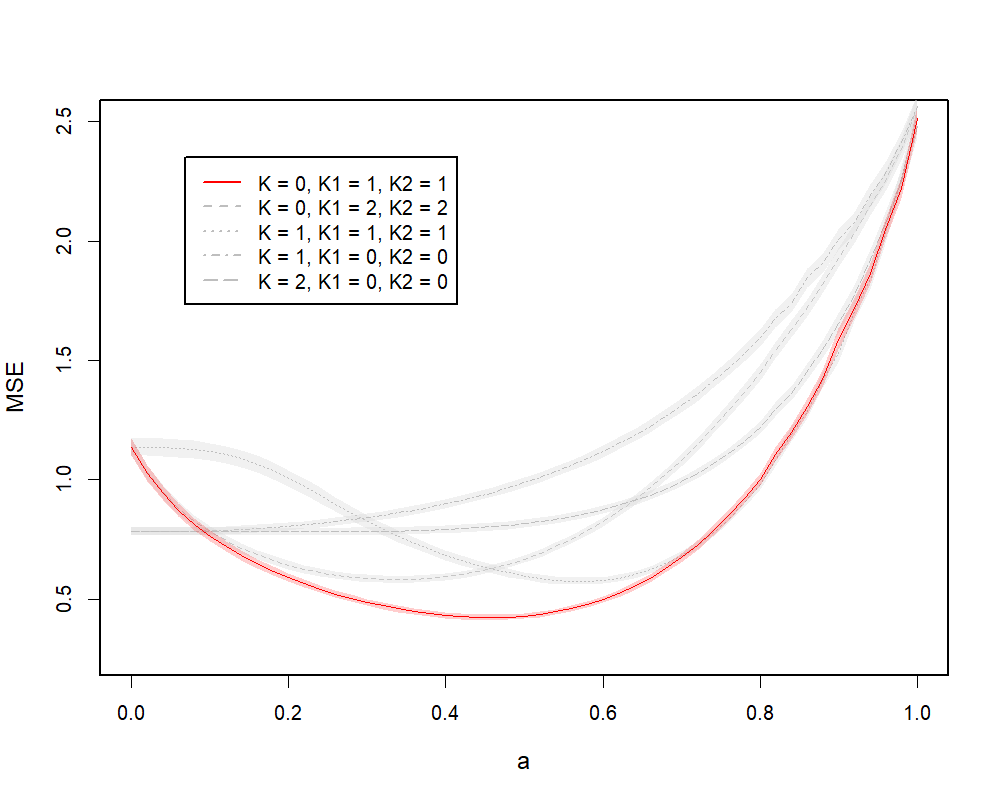}
		
		{(b) PLS Example}
	\end{minipage}
	\caption{MSE curves for JICO models with different ranks under the PCR setting (a) and PLS setting (b).  This figure appears in color in the electronic version of this article, and any mention of color refers to that version.}
	\label{fig: PCRPLSplot}
\end{figure}

We train JICO on a wide range of $\gamma \in [0, \infty)$, using different combinations of $K, K_1, K_2$, with a maximum
of $300$ iterations. Figure \ref{fig: PCRPLSplot}(a) demonstrates the MSEs evaluated on the test data over $50$ repetitions. For
better illustration, we plot MSE curves as a function of $a$, with $a = \gamma/(\gamma + 1)$, which is a one-to-one
monotone map from $\gamma \in [0, \infty)$ to $a \in [0,1]$. In particular, when $a = 0, 0.5$ and $1$, we have
$\gamma = 0, 1$ and $\infty$, which correspond to the cases of OLS, PLS and PCR respectively. The 
 solid curve
demonstrates the model performance given true ranks $K = K_1= K_2 = 1$, whereas the gray curves show the performance of
models with mis-specified ranks. In particular, we consider four mis-specified rank combinations. Among them, two rank
combinations ($K = 1$, $K_1 = K_2 = 0$; $K = 2$, $K_1 = K_2 = 0$) correspond to joint models. The other two combinations
($K = 0$, $K_1 = K_2 = 1$; $K = 0$, $K_1 = K_2 = 2$) correspond to group-specific models. We can see from Figure \ref{fig: PCRPLSplot}(a)  that the absolute minimum is given by the model with true ranks and $a = 1$, which refers to the underlying
true model. When we look at the curves on the spectrum of $a$ as a whole, the joint models with $K = 1$ or $2$,
$K_1 = K_2 = 0$ always perform worse than the model with $K = K_1 = K_2 = 1$, because they are unable to capture the
group-specific information from the underlying model. The model with true ranks performs better than the individual models
with $K = 0$, $K_1 = K_2 = 1$ or $2$ for larger values of $a$, because the latter models cannot capture as much global
information as the former. However, the model with $K = K_1 = K_2 = 1$ performs worse than the individual models for
smaller values of $a$, where the latter achieves much more acceptable performances. This means that the choice of optimal
ranks for our model can be sensitive to the choice of $\gamma$. For smaller $\gamma$ values, individual models tend to be
more reliable under the PCR setting. We notice that the end of the curve is not very smooth when $K = 2, K_1= K_2 =0$. One
possible reason is that the solution path of CR can sometimes be discontinuous with respect to $\gamma$
\citep{bjorkstrom1996continuum}, consequently the CR algorithm may be numerically unstable for certain $\gamma$ values.

We further illustrate the performance of JICO by comparing it with several existing methods. In particular, we include
ridge regression (Ridge), partial least squares (PLS) and principal component regression (PCR).
For JICO, we select the models trained under true ranks $K = K_1 = K_2 = 1$ (performance as illustrated by the
solid curve in Figure \ref{fig: PCRPLSplot}(a), with $\gamma = 0, 1, \infty$, which correspond to the cases of OLS, PLS and PCR
respectively. For a fair comparison, for PLS and PCR methods, we fix the number of components to be $2$ for both a global
fit and a group-specific fit. 
Table \ref{tbl: PCRPLSsim}(a) summarizes the MSEs and of these methods. The numbers provided in the brackets represent the standard error. The first two columns summarize the performance
for each group ($g = 1, 2$), and the last column summarizes the overall performance. The JICO model with $\gamma = \infty$
performs significantly better than the rest, because it agrees with the underlying true model. Among other mis-specified
methods, group-specific PLS is relatively more robust to model mis-specification. 

\begin{table}
		\centering
		\resizebox{\textwidth}{!}{
			\renewcommand{\arraystretch}{1.1}
			\begin{tabular}{l|llccc}
				\hline \hline
				& \multicolumn{2}{c}{Method} & $g = 1$ & $g = 2$ & Overall \\
				\hline \hline
				\multirow{9}{*}{(a) PCR Example}&\multirow{3}{*}{JICO} & $\gamma = 0$  & 1.994(0.063) & 2.012(0.068) & 2.003(0.053) \\
				& & $\gamma = 1$     & 0.679(0.018) & 0.701(0.026) & 0.69(0.017)  \\
				& & $\gamma = \infty$   &\textbf{0.04}(0.001) & \textbf{0.04}(0.001) & \textbf{0.04}(0.001)  \\
				\cline{2-6}
				& \multirow{3}{*}{Global} & Ridge &1.734(0.056) & 1.78(0.065) & 1.757(0.052)  \\
				& & PLS   & 1.163(0.033) & 1.194(0.045) & 1.178(0.031) \\
				& & PCR   & 0.946(0.04) & 0.977(0.044) & 0.961(0.022) \\
				\cline{2-6}
				& \multirow{3}{*}{Group-specific}  & Ridge & 0.252(0.009) & 0.27(0.009) & 0.261(0.005) \\
				& & PLS   & 0.254(0.009) & 0.272(0.009) & 0.263(0.005)\\
				& & PCR   & 0.68(0.042) & 0.71(0.05) & 0.695(0.032) \\
				\hline
				\multirow{9}{*}{(b) PLS Example}&\multirow{3}{*}{JICO} & $\gamma = 0$  & 0.57(0.023) & 0.567(0.021) & 0.569(0.018) \\
				& & $\gamma = 1$     & \textbf{0.211}(0.008) & \textbf{0.218}(0.008) & \textbf{0.215}(0.006) \\
				& & $\gamma = \infty$  & 1.236(0.038) & 1.277(0.037) & 1.256(0.025) \\
				\cline{2-6}
				& \multirow{3}{*}{Global} & Ridge & 1.698(0.064) & 1.742(0.06) & 1.72(0.055) \\
				& & PLS  & 0.3(0.011) & 0.297(0.01) & 0.299(0.008) \\
				& & PCR  & 1.229(0.036) & 1.298(0.041) & 1.263(0.025) \\
				\cline{2-6}
				& \multirow{3}{*}{Group-specific}  & Ridge & 0.375(0.013) & 0.425(0.016) & 0.4(0.01)\\
				& & PLS  & 0.412(0.014) & 0.406(0.016) & 0.409(0.008)\\
				& & PCR  & 1.234(0.037) & 1.25(0.036) & 1.242(0.024)  \\
				\hline
				\multirow{9}{*}{(c) OLS Example (a)}&\multirow{3}{*}{JICO} & $\gamma = 0$  & \textbf{0.082} (0.002) & \textbf{0.083 }(0.003) & \textbf{0.082} (0.002) \\
				& & $\gamma = 1$    & 0.403 (0.011) & 0.419 (0.011) & 0.411 (0.007) \\
				& & $\gamma = \infty$  & 1.006 (0.031) & 1.07 (0.03) & 1.038 (0.02) \\
				\cline{2-6}
				& \multirow{3}{*}{Global} & Ridge & \textbf{0.084} (0.004) & \textbf{0.084} (0.003) & \textbf{0.084} (0.003) \\
				& & PLS   & 0.221 (0.007) & 0.226 (0.006) & 0.223 (0.005) \\
				& & PCR   & 0.991 (0.032) & 1.069 (0.030) & 1.030 (0.020)  \\
				\cline{2-6}
				& \multirow{3}{*}{Group-specific}  & Ridge & 0.574 (0.017) & 0.599 (0.024) & 0.586 (0.013) \\
				& & PLS   & 0.572 (0.016) & 0.599 (0.024) & 0.585 (0.013) \\
				& & PCR   & 0.996 (0.032) & 1.061 (0.030) & 1.028 (0.021)  \\
				\hline
				\multirow{9}{*}{(d) OLS Example (b)}&\multirow{3}{*}{JICO} & $\gamma = 0$  &\textbf{0.063}(0.002) & \textbf{0.066}(0.004) & \textbf{0.064}(0.002) \\
				& & $\gamma = 1$     &
				0.257(0.009) & 0.27(0.009) & 0.264(0.006) \\
				& & $\gamma = \infty$   & 
				1.004(0.031) & 1.002(0.03) & 1.003(0.023) \\
				\cline{2-6}
				& \multirow{3}{*}{Global} & Ridge & 
				0.646(0.021) & 0.673(0.024) & 0.66(0.019) \\
				& & PLS    & 
				0.957(0.027) & 0.971(0.032) & 0.964(0.022) \\
				& & PCR   & 
				1.023(0.031) & 1.016(0.031) & 1.019(0.023) \\
				\cline{2-6}
				& \multirow{3}{*}{Group-specific}  & Ridge & 
				0.076(0.004) & 0.072(0.005) & 0.074(0.003)\\
				& & PLS   & 
				0.113(0.003) & 0.116(0.005) & 0.115(0.003) \\
				& & PCR   & 
				0.978(0.03) & 0.987(0.029) & 0.983(0.022) \\
				\hline\hline
		\end{tabular}
	}
	\caption{Groupwise and overall MSEs under the PCR, PLS and OLS settings. Numbers in brackets are standard errors.}
		\label{tbl: PCRPLSsim}
\end{table}

\subsection{PLS Setting}\label{sec: PLSsim}
In this section, we consider the model setup that is more favorable to $\gamma = 1$. In this scenario, the CR solutions to \eqref{eq: continuumopt} and \eqref{eq: continuumoptgroup} coincide with the PLS solutions. 
Same as in Section \ref{sec: PCRsim}, we still consider the construction of weights as linear transformations of the
eigenvectors.

Given training data $\Xb = [\Xb_1', \Xb_2']'$, denote $\Vb_{p \times q}$ as the matrix of top $q$ eigenvectors of
$\Xb'\Xb$. We let $\wb = \Vb \mathbf{1}_{q}/\sqrt{q}$, where $\bm{1}_q$ denotes a $q \times 1$ vector with elements all
equal to $1$. In this way, the $q$ top eigenvectors contribute equally to the construction of $\bS_g$. Similarly, we let
$\tilde{\Xb}_g = \Xg(\Ib - \wb \wb')$ 
and $\Vb_g$ be the $p \times q_g$ matrix of top $q_g$ eigenvectors of $\tilde{\Xb}_g'\tilde{\Xb}_g$. Then we let
$\wb_g = \Vb_g \bm{1}_{q_g}/\sqrt{q_g}$. 
To construct a model more favorable to PLS, in this section, we let $q = n/2$ and $q_g = n_g/2$. We generate $\bY_g$ from
(\ref{eq: Ylatentdecompsim}) by letting $\alpha = 1$ and $\alpha_g = 0.5$.

Similar to the PCR setting, in Figure \ref{fig: PCRPLSplot}(b), we illustrate the MSE curves of JICO models with different rank
combinations on a spectrum of $a$, where $\gamma = a/(1-a)$. Again, the  solid curve represents the model with true ranks,
while the gray curves represent models with mis-specified ranks. The absolute minimum is given by the solid curve at
$a$ around $0.5$, which corresponds to the underlying true model. Moreover, the solid curve gives almost uniformly the
best performance on the spectrum of $a$ compared with the gray curves, except on a small range of $a$ close to
$0$. Hence, under the PLS setting, the optimal ranks can be less sensitive to the  choice of $\gamma$. At initial
values of $a$, the solid curve almost overlaps with the gray curve that represents the joint model with
$K = 1, K_1 = K_2 = 0$. This means that when $\gamma$ is close to $0$, the individual signals identified by the full model
with $K = K_1 = K_2 = 1$ can be ignored. Therefore, the two group-specific models that capture more individual information
give the best performance in this case. For $a$ values closer to $1$, the gray curve that represents the joint model with
$K = 2, K_1 = K_2 = 0$ is very close to the solid curve. This means that the effects of individual components
estimated by JICO tend to become more similar across groups for larger $\gamma$.

In Table \ref{tbl: PCRPLSsim}(b), we summarize the MSEs of JICO models trained with true ranks $K = K_1 = K_2 = 1$ and $\gamma = 0, 1, \infty$, along with other methods as described in Section \ref{sec: PCRsim}. JICO with $\gamma = 1$ shows the best performance among all methods, followed by the global PLS method, since the true model favors PLS and the coefficient $\alpha_g = 0.5$ for  the group-specific component is relatively small.

\subsection{OLS Setting}
In this section, we simulate the setting that favors $\gamma = 0$, which corresponds to the case of OLS in CR. It is shown in \cite{stone1990continuum} that when $\gamma = 0$, there is only one non-degenerate direction that can be constructed from the CR algorithm. Hence, under the JICO framework, the model that favors $\gamma = 0$ embraces two special cases: a global model with $K = 1$, $K_g  = 0$ and a group-specific model with $K = 0$, $K_g = 1$. 


For the two cases, we simulate $\bY_g$ with (a) $\alpha = 1$, $\alpha_g = 0$ and (b) $\alpha = 0$, $\alpha_g = 1$ respectively. The construction of $\wb$ and $\wb_g$ is the same as  that in Section \ref{sec: PLSsim} with $q = n$ and $q_g = n_g$. 

\begin{figure}
	\begin{minipage}[b]{.5\textwidth}
		\centering
		\includegraphics[scale=0.2]{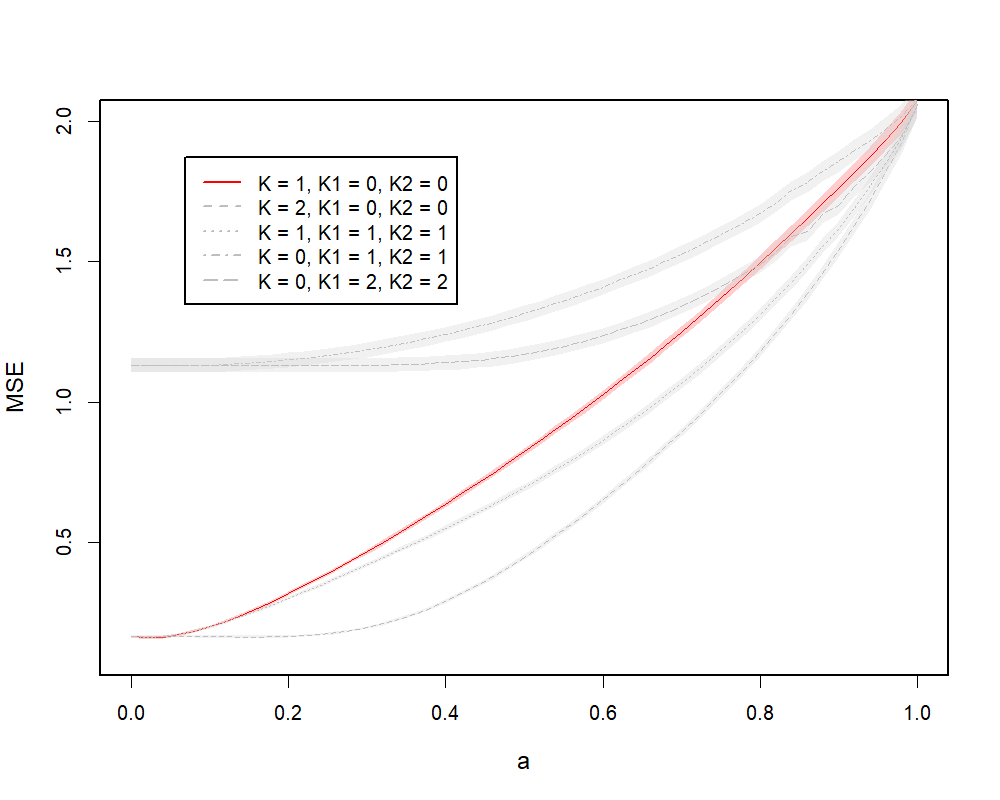}
		
		{(a) $\alpha = 1, \alpha_g = 0$}
	\end{minipage}
	\begin{minipage}[b]{.5\textwidth}
		\centering
		\includegraphics[scale=0.2]{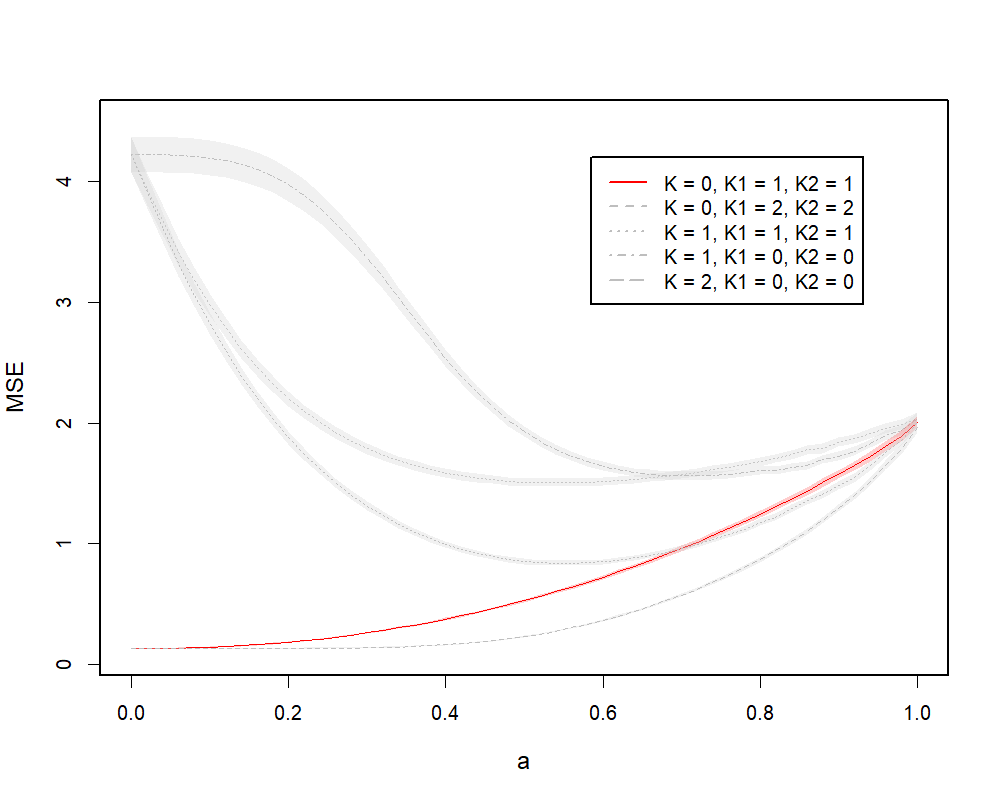}
		
		{(b) $\alpha = 0, \alpha_g = 1$}
	\end{minipage}
	\caption{MSE curves for JICO models with different ranks under OLS settings (a) and (b). (a) is generated under a global model and (b) is generated under a group-specific model. This figure appears in color in the electronic version of this article, and any mention of color refers to that version.}
	\label{fig: OLSplot}
\end{figure}

Figure \ref{fig: OLSplot} illustrates MSE curves of the two cases, where (a) represents the case of the global model and (b)
represents the case of the group-specific model. In both cases, the absolute minimum can be found on the solid curves at $a = 0$, which represents the MSE curves from the models with true ranks $K = 1, K_g = 0$ and $K = 0, K_g = 1$ respectively. In (a), there are two competitive models against the model with true ranks: another global model with $K = 2$, $K_g =0$ and the model with $K = K_1 = K_2 = 1$. They both achieve the same performance with the solid curve at $a = 0$, and stay low at larger values of $a$. This is because, when $\gamma$ is mis-specified, additional model ranks help capture more information from data. The $K = 2$, $K_g =0$ model performs better because the underlying model is a global model. This is also true for (b). The global minimum can be found at $a = 0$ on the solid curve, while the $K = 0, K_g = 2$ model performs better when $a$ gets larger. Again, this is because larger $K_g$ helps capture more information from data. The $K = K_1 = K_2 = 1$ model does not perform as well, because the estimated joint information dominates, which does not agree with the true model. We observe some discontinuities on the $K = 2, K_g = 0$ curve, since the CR solution path can sometimes be discontinuous with respect to $\gamma$ as discussed in the PCR setting in Section \ref{sec: PCRsim}.

In Table \ref{tbl: PCRPLSsim} (c) and (d), we summarize the MSEs of JICO models trained with the true ranks with $\gamma = 0, 1, \infty$ and
other methods described in Section \ref{sec: PCRsim}. For a fair comparison, the number of components for PCR and PLS is
chosen to be $1$ for both global and group-specific
fits. 
{The JICO model with $\gamma = 0$, along with Ridge always achieve better performance than all other methods. It is
  interesting to notice that in (c), the JICO models with $\gamma = 1$ and $\infty$ coincides with global PLS and PCR
  respectively, and hence they achieve the same performances. Similarly, in (d), JICO models with $\gamma = 1$ and
  $\infty$ coincide with group-specific PLS and PCR respectively, and they achieve the same performance
  correspondingly. In addition, when $K=1, K_g=0$, the solution of CR algorithm coincides with the global OLS model. Thus
  the JICO model with $\gamma=0$ and the global Ridge have similar performance in (c). Similarly, when $K=0, K_g=1$, the
  JICO model with $\gamma=0$ and group-specific Ridge have similar performance in (d).  }

\section{Applications to ADNI Data Analysis}\label{sec: jicoreal}
We apply our proposed method to analyze data from the Alzheimer’s Disease (AD) Neuroimaging Initiative (ADNI). 
 It is well known that AD accounts for most forms of dementia characterized by progressive cognitive and memory deficits. The increasing incidence of AD  makes  it a very important health issue and has attracted a lot of scientific attentions. 
  To predict the AD progression, it is very important and useful to develop efficient methods for the prediction of disease status and clinical scores (e.g., the Mini Mental State Examination
(MMSE) score and the AD Assessment Scale-Cognitive Subscale (ADAS-Cog) score). In this analysis, we are interested in predicting the ADAS-Cog score by features extracted from 93 brain regions scanned from structural magnetic resonance imaging (MRI). All subjects in our analysis are from the ADNI2 phase of the study. There are 494 subjects in total in our analysis and 3 subgroups: NC (178), eMCI (178) and AD (145), where the numbers in parentheses indicate the sample sizes for each subgroup. As a reminder, NC stands for the Normal Control, and eMCI stands for the early stage of Mild Cognitive
Impairment in AD progression. 

For each group, we randomly partition the data 
into two parts: 80\% for training the model and the rest for testing the
performance. We repeat the random split for 50 times. The testing MSEs and the corresponding
standard errors are reported in Table \ref{tbl: ADNIres}. Both groupwise and overall performance are summarized. We
compare our proposed JICO model with four methods: ridge regression (Ridge), PLS, and PCR. 
We perform both a global and a group-specific fit for Ridge, PLS, and PCR, where the regularization parameter in Ridge and
the number of components in PCR or PLS are tuned by 10-fold cross validation (CV).
For our proposed JICO model, we demonstrate the result by fitting the model with fixed $\gamma =0, 0.25, 1, \infty$, or
tuned $\gamma$ respectively. In practice, using exhaustive search to select the optimal values for $K$ and $K_g$ can be
computationally cumbersome, because the number of combinations grows exponentially with the number of candidates for each
parameter. Based on our numerical experience, we find that choosing $K_g$ to be the same does not affect the performance
on prediction too much. Details are discussed in Appendix \ref{sec:Appendix D}.  Therefore, in all these cases, the optimal ranks for
JICO are chosen by an exhaustive search in $K \in \{0,1 \}$ and $K_1=K_2=K_3\in \{0,1 \}$ to see which combination gives
the best MSE. We choose $K$ and $K_g$ to be small 
to improve our model interpretations.  The optimal value of $\gamma$ is chosen by 10-fold CV.


As shown in Table \ref{tbl: ADNIres}, JICO performs the best among all competitors. Fitting JICO with $\gamma=0.25$ yields
the smallest overall MSE. JICO with parameters chosen by CV performs slightly worse, but is still better than the other
global or group-specific methods. The results of JICO with $\gamma = 0, 1$ and $\infty$, which correspond to OLS, PLS, and
PCR, are also provided in Table \ref{tbl: ADNIres}. Even though their prediction is not the best, an interesting
observation is that they always have better performance than their global or group-specific counterparts. For example,
when $\gamma = 1$, JICO has much better overall prediction than the group-specific PLS. This indicates that it is
beneficial to capture global and individual structures for regression when subpopulations exist in the data.

In Table \ref{tbl: ADNIres}, global models perform the worst, because they do not take group heterogeneity into
consideration. The group-specific Ridge appears to be the most competitive one among group-specific methods. Note that for the AD group, our JICO model with $\gamma = 0.25$ or  tuned $\gamma$ outperforms
the group-specific Ridge method by a great margin. 

\begin{table}
	\centering
	\resizebox{\linewidth}{!}{%
	\renewcommand{\arraystretch}{1.1}
	\begin{tabular}{llllll}
		\hhline{======}	
		\multicolumn{2}{c}{Method}  & NC & eMCI & AD & Overall \\	
		\hhline{------}	
		\multirow{5}{*}{JICO} 
		& $\gamma = 0$ & 6.671 (0.137) & 11.319 (0.309) & 55.798 (1.556) & 22.821 (0.466) \\
		& $\gamma = .25$ & \textbf{6.316} (0.121) & 10.394 (0.279) & \textbf{40.853} (1.294) & \textbf{17.951} (0.394) \\
		& $\gamma = 1$  & 6.443 (0.124) & 10.353 (0.291) & 44.054 (1.449) & 18.929 (0.441) \\
		& $\gamma = \infty$  & 6.608 (0.138) & 11.121 (0.308) & 49.997 (1.832) & 21.013 (0.558) \\
		
		& CV  & 6.414 (0.129) & \textbf{10.333}(0.289) & 41.297 (1.348) & 18.096 (0.401) \\
		\hline
		\multirow{3}{*}{Global} 
		& Ridge & 23.450 (0.751) & 21.276 (0.796) & 63.989 (2.657) & 34.692 (0.840) \\
		& PLS & 26.310 (0.787) & 22.672 (0.915) & 68.193 (3.183) & 37.442 (0.982) \\
		& PCR & 25.228 (0.771) & 21.966 (0.802) & 69.541 (2.969) & 37.209 (0.907) \\
		\hline
		\multirow{3}{*}{Group-specific} 
		& Ridge & 6.336 (0.116) & 10.353 (0.278) & 42.271 (1.315) & 18.364 (0.392) \\
		& PLS & 6.656 (0.136) & 11.298 (0.306) & 48.434 (1.725) & 20.629 (0.524) \\
		& PCR & 6.656 (0.136) & 11.346 (0.304) & 47.357 (1.484) & 20.327 (0.454) \\
		\hhline{------}	
	\end{tabular}
}
	\caption{Groupwise and overall MSEs on the ADNI data. Numbers in brackets are standard errors.}
	\label{tbl: ADNIres}
\end{table}

\begin{figure}
	\begin{center}
		\begin{minipage}{1.2cm}
			\vspace{0.4cm}
			
			\ NC
			
			\vspace{5.2cm}
			
			\ AD
			
		\end{minipage}
		\begin{minipage}{8.8cm}
			\hspace{1.8cm}  $\Jb_g$ \hspace{3.5cm} $\Ab_g$
			
			\includegraphics[scale = 0.6]{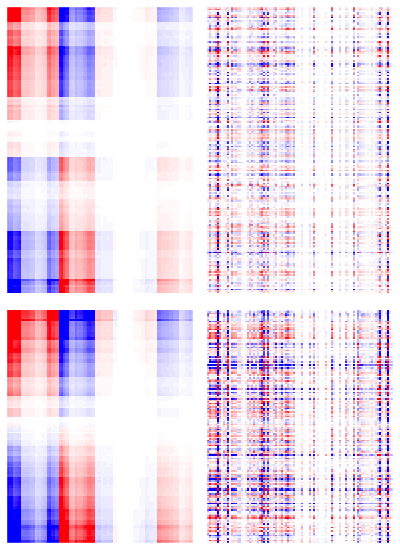}
			
			\ \ \ \ \ \ \ \ \ \ \ \ \ \includegraphics[scale=0.1]{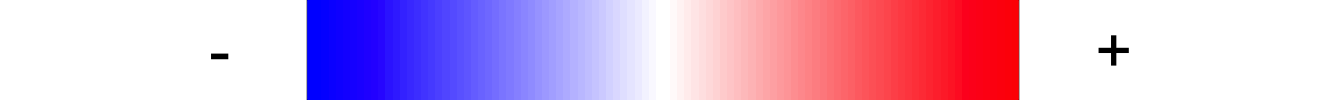}
		\end{minipage}
	\end{center}
	\caption{Heatmaps of joint and individual structures from NC and AD estimated from JICO. This figure appears in color in the electronic version of this article, and any mention of color refers to that version.}
	\label{fig: jicoheatmap}
\end{figure}

\begin{figure}
	\begin{center}
		\includegraphics[scale = 0.74]{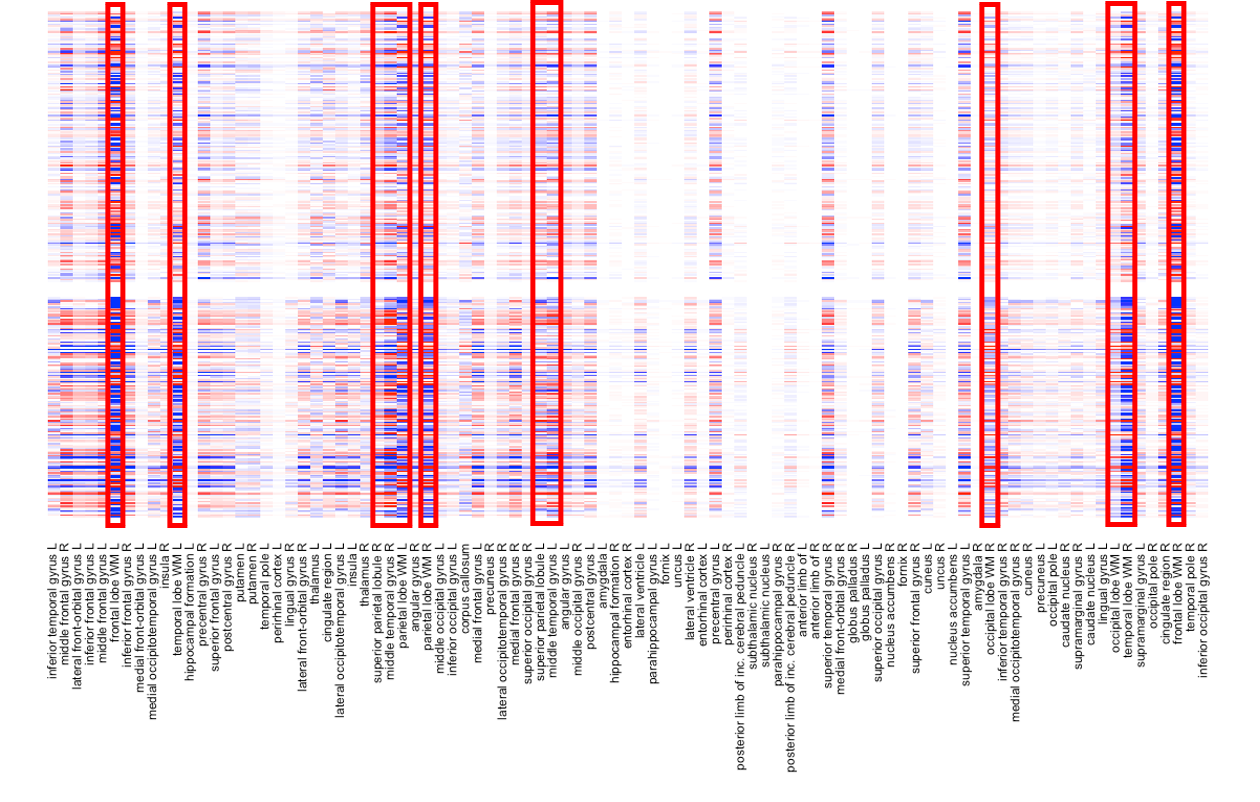}
	\end{center}
\caption{Heatmaps of individual structures from NC and AD estimated from JICO with MRI feature names. The highlighted
  regions are frontal lobe WM, temporal lobe WM, occipital lobe WM, parietal lobe WM, superior parietal lobule, middle
  temporal gyrus. The top row is from NC and the bottom row is from AD. This figure appears in color in the electronic version of this article, and any mention of color refers to that version.}
\label{fig: indjicoheatmap}
\end{figure}

To get our results more interpretable, we further apply the JICO model to NC and AD groups. We run 50 replications of
10-fold CV to see which combination of tuning parameters gives the smallest overall MSE. The best choice is
$\gamma = \infty, K = 1, K_{NC} = K_{AD} = 3$. Then, we apply JICO using this choice and tuning parameters and give the
heatmaps of the estimated $\hat{\Jb}_g$ (left column) and $\hat{\Ab}_g$ (right column) in Figure \ref{fig: jicoheatmap}.
Rows of heatmaps represent samples and columns represent MRI features. We use the Ward's linkage to perform hierarchical
clustering on the rows of $\hat{\Jb}_{g}$, and arrange the rows of $\hat{\Jb}_g$ and $\hat{\Ab}_g$ in the same order for
each group. Moreover, we apply the same clustering algorithm to the columns of $\hat{\Jb}_g$ to arrange the columns in the
same order across the two disease groups for both joint and individual structures. Figure \ref{fig: jicoheatmap} shows
that JICO separates joint and individual structures effectively. The joint structures across different disease groups
share a very similar pattern, whereas the individual structures appear to be very distinct. We further magnify the right
column of Figure \ref{fig: jicoheatmap} in Figure \ref{fig: indjicoheatmap} with the brain region names listed. 
 We find that the variation in $\hat{\Ab}_g$ for the AD group is much larger than the counterpart for the NC group. We
highlight the brain regions that differ the most between the two groups.
The highlighted regions play crucial roles in human's cognition, thus important in AD early diagnosis \citep{michon1994relation, killiany1993temporal}. 
For example,
\cite{michon1994relation} suggested that anosognosia in AD results in part from frontal
dysfunction. \cite{killiany1993temporal} showed that the temporal horn of the lateral ventricles can be used as antemortem
markers of AD. 

\section{Discussion}\label{sec: jicocon}
In this paper, we propose JICO, a latent component regression model for multi-group heterogeneous data. Our proposed
model decomposes the response into jointly shared and group-specific components, which are driven by low-rank
approximations of joint and individual structures from the predictors respectively. For model estimation, we propose an
iterative procedure to solve for model components, and utilize CR algorithm that covers OLS, PLS, and PCR as special
cases. As a result, the proposed procedure is able to extend many regression algorithms covered by CR to the setting of
heterogeneous data. Extensive simulation studies and a real data analysis on ADNI data further demonstrate the competitive
performance of JICO. 

JICO is designed to be very flexible for multi-group data. It is able to choose the optimal parameter to determine the regression algorithm that suits the data the best, so that the prediction power is guaranteed.  At the same time, it is also able to select the optimal joint and individual ranks that best describe the degree of heterogeneity residing in each subgroup. The JICO application to ADNI data has effectively illustrated that our proposed model  can provide  nice visualization on identifying joint and individual components from the entire dataset without losing much of the prediction power. 

\section*{DATA AVAILABILITY STATEMENT} The data that support the findings of this paper are available in the Alzheimer’s Disease  Neuroimaging Initiative  at \url{https://adni.loni.usc.edu/}.

\appendix
\section{Proofs}

\textbf{Proof of Lemma 1.} The proof of the first part of this lemma follows similar arguments as in \cite{feng2018angle}.  Define $\row(\Jb) = \bigcap_{g=1}^G \row(\tilde{\Xb}_g)$. We assume that $\row(\Jb) \ne \{\bm{0}\}$ for non-trivial cases. For each $g$, $\Jb_g$ and $\Ab_g$ can be obtained by projecting $\tilde{\Xb}_g$ to $\row(\Jb)$ and its orthogonal complement in $\row(\tilde{\Xb}_g)$. Thus, by construction, we have $ \tilde{\Xb}_g = \Jb_g + \Ab_g$ and $\row(\Jb_g) \perp \row (\Ab_g)$. Since for all $g$, we have $\row(\Ab_g) \perp \row(\Jb)$, then $\bigcap_{g=1}^G \row(\Ab_g)$ has a zero projection matrix. Therefore, we have $\bigcap_{g=1}^G \row(\Ab_g) = \{\mathbf{0}\}$. On the other hand, since $\row(\Jb_g) \subset \row(\tilde{\Xb}_g)$, the orthogonal projection of $\tilde{\Xb}_g$ onto $\row(\Jb_g)$ is unique. Therefore, the matrices $\Jb_g$ and $\Ab_g$ are uniquely defined. 

For the second part of Lemma 1, note that $\tilde{\bY_g} \in \col(\tilde{\Xb}_g) \subset \col(\Jb_g) + \col(\Ab_g)$. Then, $\bJ_g^Y$ and $\bA_g^Y$ can be obtained by projecting $\tilde{\bY_g}$ onto $\col(\Jb_g)$  and $\col(\Ab_g)$ respectively. Because 
$\col(\Jb_g) \perp \col(\Ab_g)$, $\bJ_g^Y$ and $\bA_g^Y$ are uniquely defined. 

\noindent \textbf{Proof of Lemma 2.}  We explicitly describe how to find $\Jb_g$, $\Ab_g$, $\Ub$, $\Ub_g$, $\Sbb_g$,
$\Tb_g$, $\alphab$, $\alphab_g$ to satisfy the requirements. First, $\Jb_g$ can be obtained by finding an arbitrary set of
bases that span $\bigcap_{g= 1}^G \row({\Xb}_g)$.  Then, $\Ab_g$ can be obtained by solving a system of linear
equations. We prove that such $\Jb_g$ and $\Ab_g$ satisfy conditions $(i)$--$(iii)$ in Lemma 1 and
$\col(\Jb_g) \perp \col(\Ab_g)$.
Given $\Jb_g$ and $\Ab_g$, we construct $\Ub, \Ub_g, \Sbb_g, \Tb_g$, so that they satisfy $\Jb_g = \Sbb_g \Ub$, $\Ab_g = \Tb_g \Ub_g$ and $\Sbb_g'\Tb_g = \bm{0}_{K\times K_g}$. 

Let $\wb_1, \ldots, \wb_K \in \bbR^p$ be an arbitrary set of bases that span $\bigcap_{g= 1}^G \row({\Xb}_g)$. Denote $\Wb_{p\times K} = [\wb_1, \ldots, \wb_K]$. Let $\Jb_g = {\Xb}_g \Wb (\Wb'\Wb)^{-1}\Wb'$.
Next, we show that $\Jb_g$ satisfies condition $(i)$ in Lemma 1 for any $0 \le g_1 \ne g_2 \le G$. 
It suffices to show that $\row(\Jb_{g_1}) \subset \row(\Jb_{g_2})$, i.e. there exists $\Cb_{g_1, n_{g_1} \times n_{g_2}}^{g_2}$, such that $\Cb^{g_2}_{g_1} \Jb_{g_2} = \Jb_{g_1}$. Since the columns of $\Wb$ form the bases of $\bigcap_{g= 1}^G \row({\Xb}_g)$, there exists $\Qb^{g_2}_{K \times n_{g_2}}$, such that $\Qb^{g_2} {\Xb}_{g_2} = \Wb'$. Then, we have $\Qb^{g_2} \Jb_{g_2} = \Qb^{g_2} {\Xb}_{g_2}\Wb (\Wb'\Wb)^{-1}\Wb' =  (\Wb'\Wb)(\Wb'\Wb)^{-1}\Wb' = \Wb'$. Let $\Rb_{g_1} = {\Xb}_{g_1} \Wb (\Wb'\Wb)^{-1}$ and $\Cb_{g_1}^{g_2} = \Rb_{g_1} \Qb^{g_2}$, then we have 
\[ \Cb_{g_1}^{g_2}\Jb_{g_2} = \Rb_{g_1} \Qb^{g_2} \Jb_{g_2} = {\Xb}_{g_1} \Wb (\Wb'\Wb)^{-1}\Wb' = \Jb_{g_1}.\]

Given $\Wb$, we propose to solve
\begin{equation}\label{eq: Ag}
\begin{bmatrix}
\Wb' \\
\Wb' {\Xb}_g' {\Xb}_g
\end{bmatrix} \xb = \bm{0},
\end{equation}
where $\xb\in \mathbb{R}^p$ is unknown.
We first show that \eqref{eq: Ag} has non-zero solutions.
Since $\Qb^{g} {\Xb}_g = \Wb'$, \eqref{eq: Ag} can be rewritten as $\tilde{\Qb}^{g} {\Bb}_{g}{\Xb}_g \xb = \bm{0},$ where
\[ \tilde{\Qb}^g =  \begin{pmatrix}
\Qb^{g} & \\
& \Qb^g
\end{pmatrix} \text{ and } {\Bb}_g = \begin{pmatrix}
\Ib_{n_g \times n_g} \\
{\Xb}_g {\Xb}_g'
\end{pmatrix}. \]
Since \[\rank(\tilde{\Qb}^g {\Bb}_g{\Xb}_g ) \le \min\{\rank(\tilde{\Qb}^g), \rank({\Bb}_g), \rank({\Xb}_g)\} \le \min\{2K, n_g, \rank(\Xb_g)\} \le \rank(\Xb_g) < p, \]
\eqref{eq: Ag} has non-zero solutions. 

Let $\Wb_g = [\wb_{g,1}, \ldots, \wb_{g,K_g}]$ be an arbitrary set of bases that spans the space of solutions to \eqref{eq: Ag}. Let $\Ab_g = {\Xb}_g\Wb_g {(\Wb_g'\Wb_g)}^{-1} \Wb_g'.$
Next we show that $\Ab_g$ satisfies $(ii)-(iii)$ in Lemma 1 and $\col(\Ab_g) \perp \col(\Jb_g)$.
Since columns of $\Wb_g$ satisfy \eqref{eq: Ag},  we have $\Wb'\Wb_g = \bm{0}$ and $\Wb'{\Xb}_g'{\Xb}_g\Wb_g = \bm{0}$. Then, by definition, 
$\Ab_g' \Jb_g = \bm{0}$ and $\Jb_g \Ab_g' = \bm{0}$, which imply that $\row(\Ab_g) \perp \row(\Jb_g)$  and 
$\col(\Ab_g) \perp \col(\Jb_g)$. 
Since for all $g$, $\row(\Ab_g)$ is perpendicular to the columns of $\Wb$, 
then we have $\bigcap_{g=1}^G \row(\Ab_g)=\{\bm{0}\}$.

Finally, letting $\Sbb_g = {\Xb}_g \Wb $, $\Tb_g = {\Xb}_g \Wb_g$, $\Ub = (\Wb'\Wb)^{-1} \Wb'$ and
$\Ub_g = (\Wb_g'\Wb_g)^{-1} \Wb_g'$,  we have $\Sbb_g'\Tb_g = \bm{0}$.
Let ${\alphab} = ({\Sbb}'{\Sbb})^{-1} {\Sbb}'\bY$ and ${\alphab}_g = ({\Tb}_g'{\Tb}_g)^{-1}{\Tb}_g'{\bY}_g$, where
$\Sbb = \Xb \Wb$. Then we obtain $\bJ^{Y}_g$ and $\bA^{Y}_g$ by letting $\bJ^{Y}_g = \Sbb_g \alphab \in \col(\Jb_g)$ and
$\bA^{Y}_g = \Tb_g \alphab_g \in \col(\Ab_g)$. 


%
%
%

\section{Derivation of the CR algorithm for solving \eqref{eq: continuumjoint} and \eqref{eq: continuumindv}}\label{sec:Appendix B}

Consider the following CR problem that covers \eqref{eq: continuumjoint} and \eqref{eq: continuumindv} as special
        cases:
\begin{equation}\label{eq: continuumunify}
\begin{aligned}
\max_\wb\quad
&\cov(\Xb\wb, \bY)^2 \var(\Xb\wb)^{\gamma-1}\\
\text{s.t.} \quad&\wb' \wb = 1, \\
&\wb' \Xb'\Xb\wb_j =0; \quad   j = 1, \ldots, k -1 \text{ if }k\ge 2, \\
&\hat{\Wb}'\wb = \mathbf{0}; \\
&\hat{\Sbb}'\Xb \wb=\mathbf{0}, \\
\end{aligned}
\end{equation}
where $\hat{\Wb}$, $\hat{\Sbb}$, $\Xb$ and $\bY$ are given a priori.

The solution to (S2) resides in $\col(\hat{\Wb})^{\perp}$, i.e. the space that is orthogonal to the columns of $\hat{\Wb}$. Hence, we let $\hat{\Pb }= \hat{\Wb}(\hat{\Wb}'\hat{\Wb})^{-1}\hat{\Wb}'$ and $\hat{\Xb} = \Xb (\Ib - \hat{\Pb})$,  the latter being the projection of $\Xb$ into the space that is orthogonal to the columns of $\hat{\Wb}$.
Let $m$ be the rank of the matrix $\hat{\Xb}$, and $\hat{\Xb} = \Ub \Db\Vb'$ be the corresponding rank-$m$ SVD decomposition, with $\Ub \in \bbR^{n \times m}$, $\Vb \in \bbR^{p \times m}$, $\Db$ the $m \times m $ diagonal matrix. 
Since the representation of the solution to (S2) might not be unique, to avoid ambiguity, we write the solution as the linear combination of the column vectors of $\Vb$, i.e. $\wb = \Vb \zb$, for some $\zb \in \bbR^m$. Note that all $\wb = \Vb \zb$ satisfies $\hat{\Wb}'\wb = \bm{0}$. Hence, the constraint $\hat{\Wb}'\wb = \bm{0}$ from (S2) can be satisfied under this representation. 

At step $k+1$, the original optimization problem can be reformulated as follows:
\begin{equation}\label{eq: continuumopt2}
\max_{\zb}\ (\zb'\db)^2 ( \zb' \tilde{\Eb} \zb)^{\gamma-1} \quad \text{s.t. } \zb'\zb = 1,\zb'\tilde{\Eb}\Zb_{k}  = \bm{0}, \text{ and } \zb'\Db \Ub'\hat{\Sbb}= \bm{0},
\end{equation}
where $\db = \Vb' \hat{\Xb}'\bY$, $\Zb_{k} = [\zb_1, \ldots, \zb_{k}]$ and $\tilde{\Eb} = \Db^2$. To solve \eqref{eq: continuumopt2}, we can expand the objective to its Lagrangian form:
\[ T^*(\zb) =  (\zb'\db)^2 ( \zb' \tilde{\Eb} \zb)^{\gamma-1} - \lambda_0 (\zb'\zb - 1) - 2\zb'[\tilde{\Eb} \Zb_k \ \Db \Ub'\hat{\Sbb}] \bm{\Lambda}_k, \]
where $\Lambdab_k = [\lambda_1, \ldots, \lambda_{k+K}]'$ and $\lambda_0, \ldots,  \lambda_{k+K}$ are Lagrange multipliers. To solve \eqref{eq: continuumopt2}, we take the derivative of $T^*$ with respect to $\zb$, then the optimizer should be the solution to the following:
\begin{equation}\label{eq: continuumderiv}
{\partial T^* \over \partial \zb }=   2(\zb' \db) ( \zb' \tilde{\Eb} \zb)^{\gamma-1} \db + 2(\gamma-1) (\zb' \db)^2( \zb' \tilde{\Eb} \zb)^{\gamma-2} \tilde{\Eb}\zb - 2\lambda_0 \zb - 2[\tilde{\Eb} \Zb_k \ \Db \Ub'\hat{\Sbb}]  \Lambdab_k = \bm{0}. 
\end{equation}
Left multiply $\zb'$ to \eqref{eq: continuumderiv} and apply the constraints 
then we can conclude that $\lambda_0 = \gamma (\zb'\db)^2 ( \zb' \tilde{\Eb} \zb )^{\gamma-1}$.
Plug this back to \eqref{eq: continuumderiv}, and let $\tau = \zb'\db$ and $\rho = \zb'\tilde{\Eb}\zb $, then we have
\[ \gamma\tau^2 \rho^{\gamma-1}\zb+ (1-\gamma) \tau^2\rho^{\gamma - 2} \tilde{\Eb}\zb +[\tilde{\Eb} \Zb_k \ \Db \Ub'\hat{\Sbb}] \Lambdab_k = \tau\rho^{\gamma-1}\db. \]
A simple reformulation of the above plus the constraints 
$\zb'[\tilde{\Eb} \Zb_k \ \Db \Ub'\hat{\Sbb}]  = \bm{0}$ yields the following matrix form:
\begin{equation}\label{eq: continuummtx}
\left[\begin{array}{cc}
\Ab& \Bb\\
\Bb' & \bm{0}
\end{array}\right]
\left[\begin{array}{c}
\zb \\
\Lambdab_k
\end{array}\right] = \left[ \begin{array}{c}
\qb\\ \bm{0} 
\end{array}\right],
\end{equation}
where $\Ab = \tau^2[\gamma \rho^{\gamma-1}\Ib + (1-\gamma) \rho^{\gamma - 2} \tilde{\Eb}]$, $\Bb = [\tilde{\Eb} \Zb_k \ \Db \Ub'\hat{\Sbb}] $, and $\qb = \tau\rho^{\gamma-1}\db$. By the standard formula for inverse of a partitioned matrix and the constraint that $\zb'\zb = 1$, we obtain
\begin{equation}\label{eq: continuumz}
\zb = {\Mb\qb\over  \|\Mb\qb\|},
\end{equation}
where $\Mb = \Ab^{-1} - \Ab^{-1} \Bb (\Bb'\Ab^{-1}\Bb)^{-1}\Bb'\Ab^{-1}$.  Note that $\Mb$ has a factor of $\tau^{-2}$ and
$\qb$ has a factor of $\tau$, hence $\Mb \qb$ has a factor of $\tau^{-1}$ that gets canceled out during the normalization
in \eqref{eq: continuumz}. Therefore, $\zb$ does not rely on the quantity of $\tau$. Without loss of generality, we can
choose $\tau = 1$. Then the only unknown parameter is $\rho$. Hence, we can formulate \eqref{eq: continuummtx} and \eqref{eq:
  continuumz} as a fixed point problem of $\zb(\rho)'\tilde{\Eb}\zb(\rho)$ as a function of $\rho$. More specifically, we
seek for $\rho^*$ that satisfies $\zb(\rho^*)'\tilde{\Eb}\zb(\rho^*) = \rho^*$. Afterwards, we obtain
$\zb^* = \Mb^*\qb^*/\|\Mb^*\qb^*\|$, where $\Mb^*$ and $\qb^*$ are computed from $\rho^*$. We summarize the procedure to
solve \eqref{eq: continuumopt2} as in the following two steps:

\begin{minipage}{0.96\textwidth }
	\vspace{3mm}
	
	\textbf{Step 1}: Solve the fixed point $\rho^*$ for $\rho = \zb(\rho)'\tilde{\Eb}\zb(\rho)$ with $\zb(\rho) = {\Mb\qb/  \|\Mb\qb\|}$, where
	\begin{align*}
	\Mb &= \Ab^{-1} - \Ab^{-1} \Bb (\Bb'\Ab^{-1}\Bb)^{-1}\Bb'\Ab^{-1},\\
	\Ab  &= \gamma \rho^{\gamma-1}\Ib + (1-\gamma) \rho^{\gamma - 2} \tilde{\Eb}, \\
	\Bb &= [\tilde{\Eb} \Zb_k \ \Db \Ub'\hat{\Sbb}],\\
	\qb &= \rho^{\gamma-1}\db;
	\end{align*}
	
	\textbf{Step 2}:  Compute $\Mb^*$ and $\qb^*$ from the fixed point $\rho^*$ in Step 2. The solution to \eqref{eq: continuumopt2} is then given by $\zb^* = \Mb^*\qb^*/\|\Mb^*\qb^*\|$.
	\vspace{3mm}
\end{minipage}

\noindent The most challenging step is Step 2, where a nonlinear equation needs to be solved. This can be done numerically by several existing algorithms. For example, we can use Newton's method \citep{kelley2003solving}, which is implemented by many optimization packages and gives fast convergence in practice.

\section{JICO iterative algorithm}\label{sec:Appendix C}
After each iteration, we obtain $K$ joint objective values as a $K \times 1$ vector:
\begin{equation}\label{eq: jointobj}
\mathcal{L}= \diag(\Wb'\Xjoint{'}\Xjoint\Wb)^{\gamma - 1}(\Wb'\Xjoint{'}\Yjoint)^2, 
\end{equation}
and $K_g$ individual objective values as a $K_g \times 1$ vector for $g  =1, \ldots, G$:
\begin{equation}\label{eq: individualobj}
\mathcal{L}_g =  \diag(\Wb_g'\Xindv{'}\Xindv\Wb_g)^{\gamma - 1}(\Wb_g'\Xindv{'}\Yindv)^2,
\end{equation} 
and compare them with the corresponding vectors obtained from the previous iteration step. Our iterative procedure stops until the differences between two consecutive iteration steps are under a certain tolerance level. Empirically, the algorithm always
met convergence criteria, albeit there are no theoretical guarantees for it. 

\begin{algorithm}[h]
	\KwData{$\{\Xb_g, \bY_g\}_{g = 1}^G$;} \textbf{Parameters:} tolerance level $\tau$; joint rank $K$ and individual rank $K_g$\;
	Initialize $\Tb_g = \bm{0}_{n_g \times K_g}$; $\Ug = \bm{0}_{K_g \times p}$; $\alphab_g = \bm{0}_{K_g \times 1}$; 
	\;
	\While{Euclidian distance $\|\nabla\mathcal{L}\|, \|\nabla\mathcal{L}_g\| > \tau$}{
		$\Xgjoint= \Xb_g - \Tb_g \Ub_g $; $\bY_g^{\text{Joint}} = \bY_g - \Tb_g \alphab_g$\;
		Estimate $\Wb = (\wb_1, \ldots, \wb_K)$ from \eqref{eq: continuumopt}\; 
		$\Sbb = \Xjoint\Wb$; $\Ub = (\Wb'\Wb)^{-1}\Wb'$; $\alphab = (\Sbb'\Sbb)^{-1} \Sbb' \Yjoint$\; 
		\For{$g = 1, \ldots, G$}{
			$\Xindv = \Xb_g - \Sbb_g \Ub$; $\Yindv = \bY_g - \Sbb_g\alphab$\;
			Estimate $\Wb_g = (\wb_{g1}, \ldots, \wb_{gK_g})$ from \eqref{eq: continuumoptgroup}\; 
			$\Tb_g = \Xindv \Wb_g$; $\Ub_g = (\Wb_g'\Wb_g)^{-1}\Wb_g'$; $\alphab_g = (\Tb_g'\Tb_g)^{-1} \Tb_g' \Yindv$; 
		}
		$\tWb = [\Wb_1, \ldots, \Wb_G]$\; 
	}
	\caption{JICO Algorithm}
	\label{alg: JICO}
\end{algorithm}

\section{Selection of Tuning Parameters}\label{sec:Appendix D}
To apply the proposed JICO method, we need to select tuning parameters $K$, $K_g$ and $\gamma$.  Based on our numerical
experience, we find that the performance of JICO depends more on the choice of $K$ but less on $K_g$. We also find that
even when the true $K_g$ differs in multiple groups, choosing them to be the same doesn't affect the MSE in predicting the
response too much. Thus, to saving computational time, we can choose $K_g$ to be the same for all groups. We propose to do
an exhaustive search for $0\leq K\leq D_1$ and $0\leq K_g\leq D_2$ to find their best combination that gives the smallest
MSE, where $D_1$ and $D_2$ are two user-defined constants. In this way, we need to try $(D_1+1)\times (D_2+1)$
combinations in total. Since we assume $\mathbf{J_g}$ and $\mathbf{A}_g$ are both low-rank, we can set $D_1$ and $D_2$ to
be relatively small. As for the tuning of $\gamma$, we propose to perform a grid search and choose the one that can
minimize the MSE in predicting the responses. 
%

\section{The impact of using different initial value $\mathbf{I}_{g}$}\label{sec:Appendix E}
The following simulation study illustrates the impact of using different matrices as the initial value for our algorithm.

We use the same data generating scheme as described in Section 4. We fix $G=2, p=200, n_{1}=n_{2}=50$. In each replication, we generate 100 training samples to train the models and evaluate the corresponding Mean Squared Error (MSE) in an independent test set of 100 samples. We also record the objective values as described below:
\begin{itemize}
	\item When $\gamma=0$, we compare the converged objective value 
	as in \eqref{eq: continuumindv}
	for $g=1, 2$. 
	\item When $\gamma=1$, we compare $\cov(\Xb^{\text{Joint}}\wb, \bY^{\text{Joint}})^2$ and $\cov(\Xindv\wb, \Yindv)^2$ for $g=1, 2$.
	\item When $\gamma=\infty$, we compare $\var(\Xb^{\text{Joint}}\wb)$ and $\var(\Xindv\wb)$ for $g=1, 2$.\
\end{itemize}

\begin{figure}
	\centering
	\includegraphics[width=\linewidth]{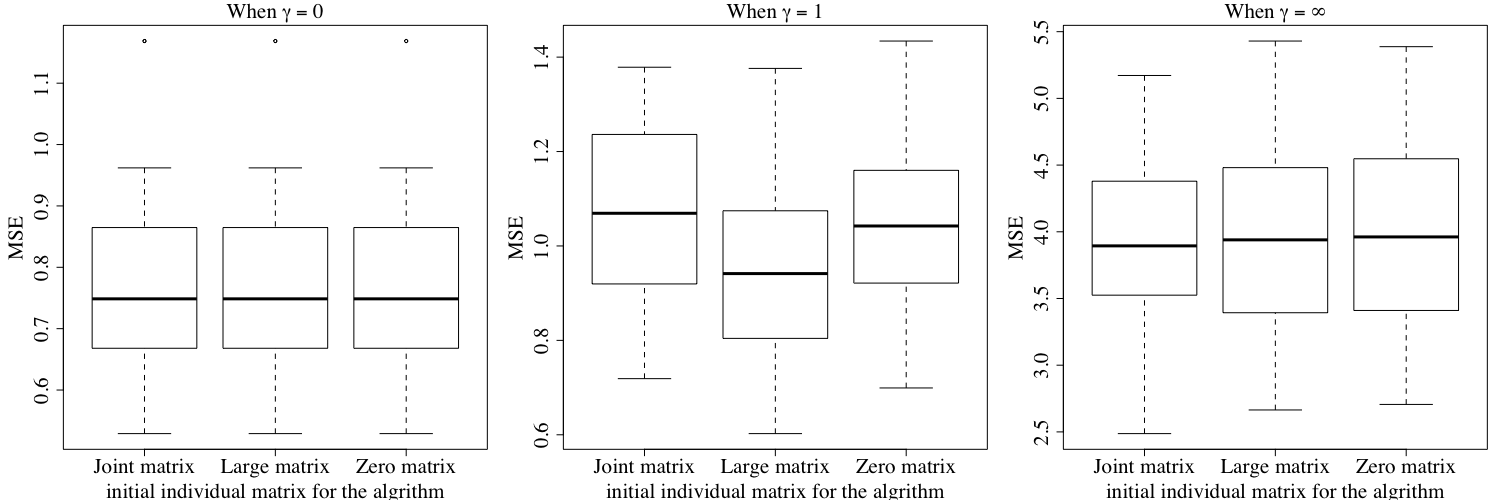}
	\caption{The impact of using different initial individual matrix on MSE under different settings.}
	\label{fig:initialcompmse}
\end{figure}

\begin{figure}
\centering
\includegraphics[width=\linewidth]{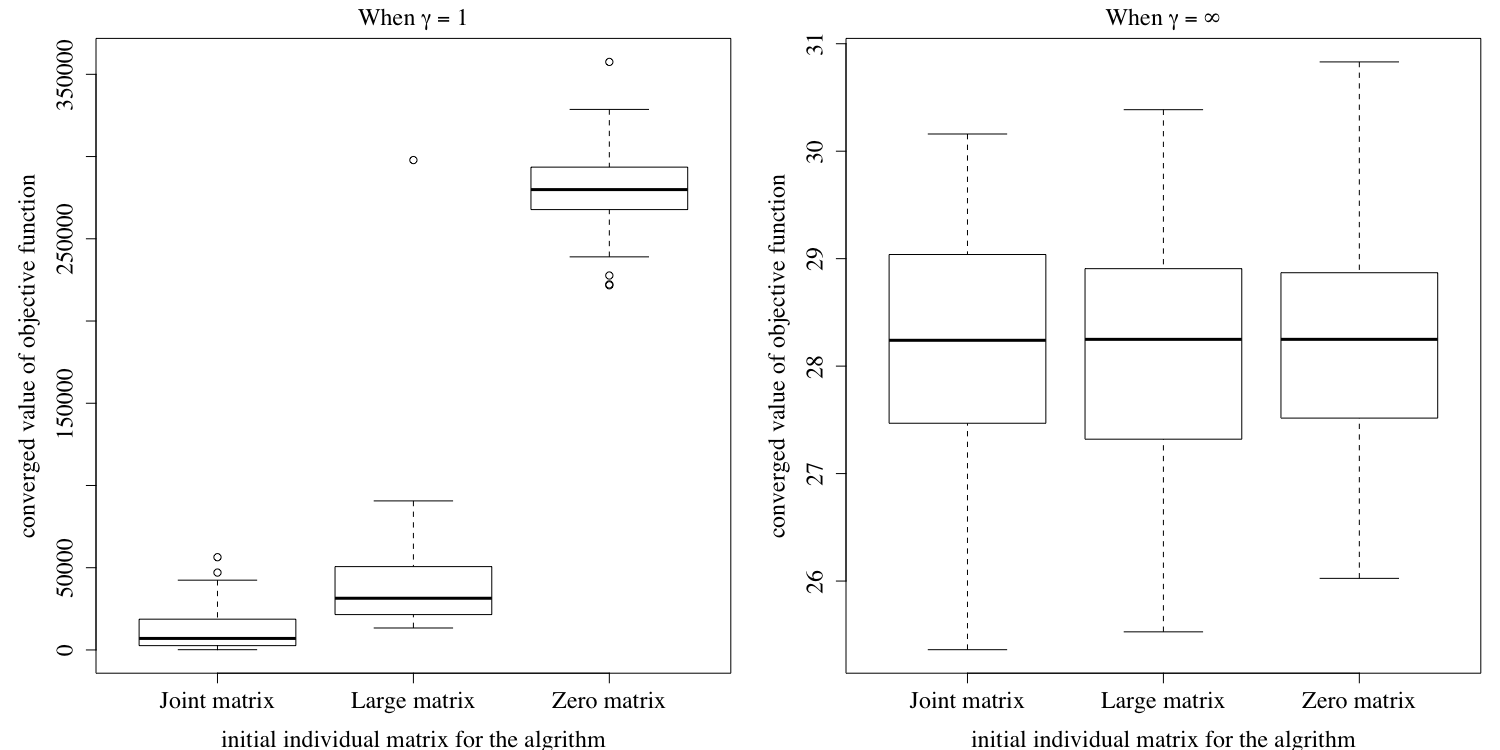}
\caption{The impact of using different initial individual matrix on converged value of the joint objective function under different settings.}
\label{fig:initialcompobjhomo}
\end{figure}

\begin{figure}
	\centering
	\includegraphics[width=\linewidth]{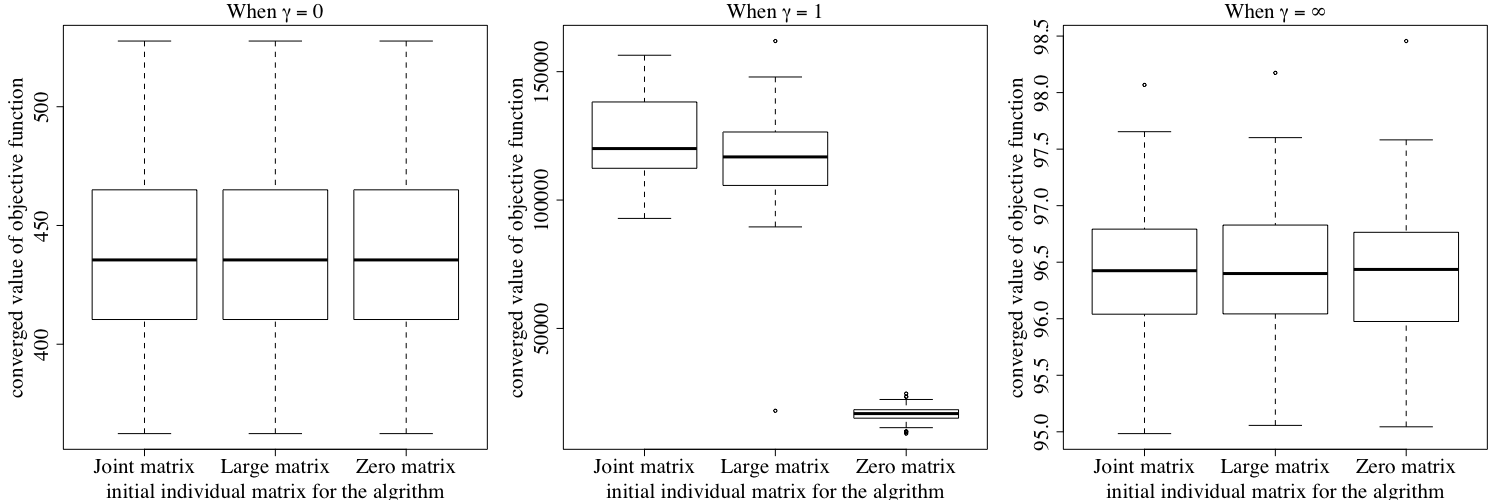}
	\caption{The impact of using different initial individual matrix on converged value of the first individual objective function under different settings.}
	\label{fig:initialcompobjheter1}
\end{figure}

\begin{figure}
	\centering
	\includegraphics[width=\linewidth]{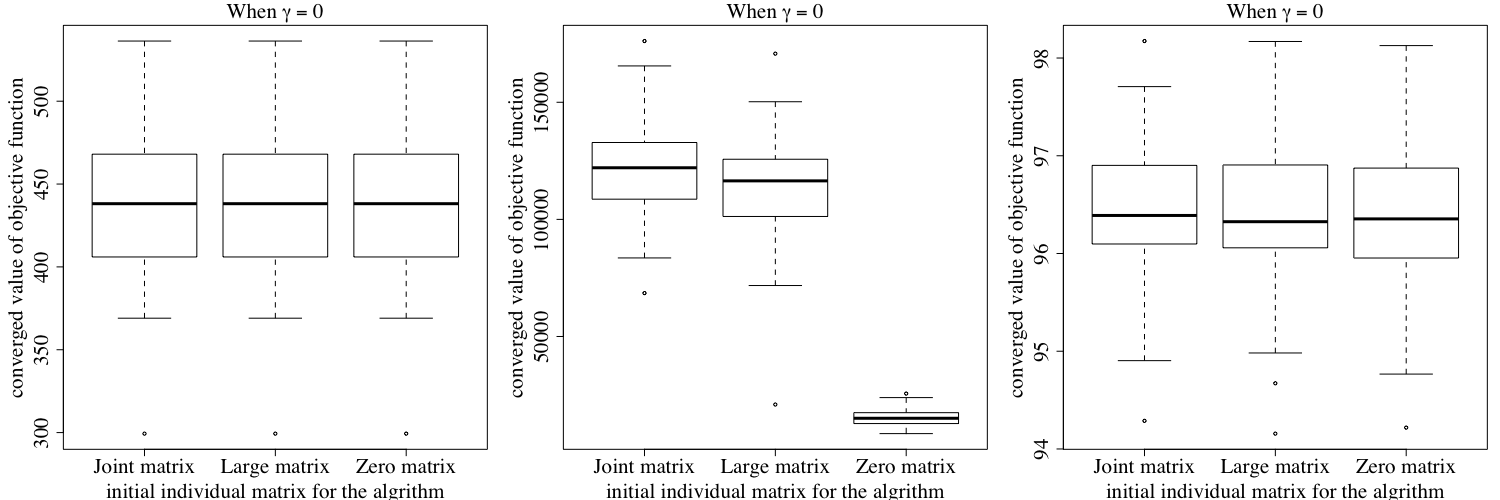}
	\caption{The impact of using different initial individual matrix on converged value of the second individual objective function under different settings.}
	\label{fig:initialcompobjheter2}
\end{figure}

We compare the performance of our algorithm by giving the boxplots of the MSEs and the converged objective values using
different initial values. We repeat the simulation for 50 times, with three kinds of initial values of the individual
matrices $\mathbf{I}_{g}$ as follows:

\begin{itemize}
	\item Zero matrix: In this setting, our algorithm starts with zero matrices as the initial value, where all entries in $\mathbf{I}_{g}$ are equal to 0. 
	\item Large matrix: In this setting, our algorithm starts with matrices with large entries as the initial value, where all entries in $\mathbf{I}_{g}$ are equal to 3. 
	\item Joint matrix: In this setting, our algorithm starts with estimated joint matrices $\mathbf{\hat{J}}_{g}$ as the initial value, where $\mathbf{\hat{J}}_{g}$ are derived from our first setting, that is using zero matrices as the initial value. 
\end{itemize} 
Figures \ref{fig:initialcompobjhomo} shows the converged value of the joint objective function under the three different initial values of the individual matrices $\mathbf I_g$ mentioned above. Similarly, Figure \ref{fig:initialcompobjheter1}  shows the converged value of the first  objective function under the three different initial values. Finally, Figure \ref{fig:initialcompobjheter2} shows the the converged value of the second individual objective function under three different initial values. The results show that our algorithm may not converge to the same objective function value using different initial values when $\gamma=1$, but Figure \ref{fig:initialcompmse} shows that our results tend to have a similar performance in terms of prediction error, no matter how we choose the initial individual matrices for our algorithm. To ensure a better performance, we recommend running our algorithm on the same dataset multiple times with different initial values, and then choose the result with the best performance. 

\begingroup
\setstretch{1}
\bibliographystyle{apalike} 
\bibliography{references}
\endgroup
\end{document}